\documentclass[%
 aip,
 amsmath,amssymb,
reprint, onecolumn %%%%% Comment onecolumn to override one column format %%%%%%%
]{revtex4-1}

\usepackage{graphicx}% Include figure files
\usepackage{dcolumn}% Align table columns on decimal point
\usepackage{bm}% bold math
\usepackage[utf8]{inputenc}
\usepackage[T1]{fontenc}
\usepackage{mathptmx}
\usepackage{etoolbox}
\usepackage{hyperref}
\usepackage{hyperref}

\hypersetup{
    colorlinks=true,
    linkcolor=blue,
    citecolor=red,
    urlcolor=blue,
    }

\makeatletter
\def\@email#1#2{%
 \endgroup
 \patchcmd{\titleblock@produce}
  {\frontmatter@RRAPformat}
  {\frontmatter@RRAPformat{\produce@RRAP{*#1\href{mailto:#2}{#2}}}\frontmatter@RRAPformat}
  {}{}
}%
\makeatother

\usepackage{amsmath,amssymb,graphicx}
\usepackage{amsthm}
\usepackage{derivative}

\newtheorem{thm}{Theorem}[section]

\newtheorem{prop}[thm]{Proposition}
\theoremstyle{definition}

\theoremstyle{remark}

\renewcommand{\thesection}{\arabic{section}}
\renewcommand{\thesubsection}{\arabic{section}.\arabic{subsection}}
\renewcommand{\thesubsubsection}{\arabic{section}.\arabic{subsection}.\arabic{subsubsection}}

\numberwithin{equation}{section}

\begin{document}

%\preprint{AIP/123-QED}

\title[]{On Classification and  Geometric Characterizations of Ensembled $2\times2$ Pseudo Hermitian and PT-Symmetric Matrices}
\author{Stalin Abraham}
\email{stalinabraham.v@gmail.com}
 \affiliation{ 
School of Physical Sciences, UM-DAE Centre for Excellence in Basic Sciences, University of Mumbai, Kalina Campus, Mumbai 400 098, India
}%

\author{Ameeya A. Bhagwat}%
 \affiliation{ 
School of Physical Sciences, UM-DAE Centre for Excellence in Basic Sciences, University of Mumbai, Kalina Campus, Mumbai 400 098, India
}%

\date{\today}% It is always \today, today,

\begin{abstract}
\textbf{Abstract}
Non-Hermitian matrices $H\in M_2(\mathbb{C})$  satisfying the 
relation $ H^{\dag}G = GH $, for invertible and singular Hermitian matrices $G$ have been studied. The matrices $H$ corresponding to invertible $G$ are known in the literature as G-pseudo Hermitian matrices. We label the matrices corresponding to the singular $G_s$ as $G_s$-pseudo Hermitian. We have proved that all $ 2\times 2$ $G$-pseudo Hermitian matrices  are PT-symmetric. For a given $G$ ($G_s$), all  $G$ ($G_s$)-pseudo-Hermitian $H\in M_2(\mathbb{C})$  are found  to be expressed as a linear variety. It is further found that for any two Hermitian $G_i,G_j\in M_2(\mathbb{C})$ such that $G_i\neq \lambda G_j$, there always exists exactly one trace less $H\in M_2(\mathbb{C})$ (up to real scaling) which is pseudo-Hermitian with respect to both these $G$ matrices. The set of all $G$- and $G_s$- pseudo-Hermitian matrices has been divided into seven distinct ensembles of matrices and the set of all PT-symmetric matrices in $M_2(\mathbb{C})$ is partitioned into four cells, denoted by $S_1,S_2,S_3$ and $S_4$. The ensembles of  trace-less G-pseudo Hermitian matrices are shown to be written as a linear combination of three basis elements from these cells. When $\mathrm{Tr}(G) = 0$, one basis element is from  $S_1$ and the other two are from $S_2$. On the other hand, when $\mathrm{Tr}(G)\neq0$, one basis element is from $S_1$ and the other two are from $S_4$. The determinant of such ensembles of trace-less matrices are shown to be quadrics, which could be hyperboloid of two sheets, hyperboloid of one sheet, ellipsoid or quadric cone for invertible $G$, whereas it is two parallel planes or a plane for singular $G_s$. Finally, the set of all the matrices $G\in M_2(\mathbb{C})$, satisfying $H^{\dagger}G = GH$, given a specific $H\in M_2(\mathbb{C})$, are shown to be describable in terms of quadratic variety.

\end{abstract}

\maketitle

\section{Introduction}
Physically measurable quantities or observables are postulated to be Hermitian operators on a suitable Hilbert space in quantum mechanics. When measurements of such observables are carried out, one obtains the eigenvalues of the operators as the measured quantities. Such measured quntities are usually expected to be real. Thus the reality of eigenvalues and the operators that are guaranteed to possess real eigenvalues have great significance in quantum mechanics. Quite surprisingly, in late 90s Bender and collaborators found that there exist non-Hermitian  operators which exhibit parity - time reversal (PT) symmetry can posses
 real spectrum \cite{Bender:1998ke,Bender:1998gh,Bender:2002vv}. This observation leads to the conclusion that Hermiticity is sufficient to ensure the reality 
of eigenvalues but it is not the necessary condition for it. 

 Non-Hermitian operators with PT-symmetry may possess complex as well as real eigenvalues. If all the eigenstates of a PT-symmetric operator are also eigenstates of a PT-operator, then all of its eigenvalues are real. Otherwise,  the PT-symmetric operator has at least one complex conjugate pairs of eigenvalues \cite{Bender:2003gu}. The first case is referred as unbroken PT-symmetry and the later is referred as broken PT-symmetry \cite{Bender:2003gu}. In early 2000's, Ali Mostafazadeh showed that PT-symmetric operators belong to a  class of non-Hermitian operators  known as pseudo-Hermitian operators and had initiated the investigations on relation between them\cite{Mostafazadeh:2001jk,Mostafazadeh:2001nr}.  Necessary and sufficient conditions for PT-symmetric and non- Hermitian Hamiltonians to have real spectra has been explored by Bender and Mostafazadeh \cite{Mostafazadeh:2001nr,Bender:2009mq}. 

Mostafazadeh proved that all diagonalizable PT-symmetric Hamiltonians are pseudo-Hermitian \cite{Mostafazadeh:2002id}. In 2019, Ruili Zhang and collaborators have shown that in finite-dimensional cases, a PT-symmetric Hamiltonian is necessarily pseudo-Hermitian regardless of its diagonalisability \cite{Zhang:2019gyc}. 
Further, the equivalence of PT-symmetric and pseudo-Hermitian matrices for 
two-level systems was shown by Qing-hai Wang  \cite{Wang201322}. 
Over the last two decades several detailed studies have been published on non-Hermitian Hamiltonians with PT-symmetry and their applications\cite{ruter2010observation,Schindler:2011mnw,Bender:2012je,bittner2012pt,zhang2016observation,10.1063/5.0002958}.

The investigations of non-Hermitian Hamiltonians  is still an ongoing activity. Non-Hermitian operators with real (complex) eigenvalues can be the Hamiltonians of non-dissipative (dissipative) systems 
\cite{Wang201322}, which is what makes such studies 
highly interesting and relevant.
With this motivation, the present work aims to carry out a systematic
investigation of the set of all the $2\times 2$ PT-symmetric and 
pseudo-Hermitian matrices, in general, non-Hermitian matrices satisfying the equation $H^{\dag}G= GH$, for a Hermitian $G$. If $G$($G_s$) is invertible (singular) the above equation defines $G(G_s)$-pseudo Hermitian matrices. Such matrices may naturally represent the operators pertaining to the 2-level quantum mechanical systems. Note that for a given $G$, the matrix $H$ satisfying the above equation, and for a given $H$, the matrix $G$ satisfying the same equation is not unique, i.e, there could exist more than one $H$, for a given $G$ and more than one $G$ for a given $H$ satisfying the equation.

In this work, we propose
a partition of the $2\times 2$ PT-symmetric matrices in section~\ref{sec:2}, leading naturally to 
systematization of their properties, and facilitating investigations of 
the more general class of operators, namely, the $G(G_s)$-Pseudo-Hermitian operators.
 In section~\ref{sec:3}, we furnish 
an explicit proof of the result that 
in two-level systems, an operator is $G$-pseudo-Hermitian only if it is PT-symmetric . We also have shown that when $\mathrm{det}(G)>0$,   $G$-pseudo Hermitian matrix $H$ have real eigenvalues and when $\mathrm{det}(G)<0$, $G$-pseudo Hermitian matrix $H$ can have real spectrum or complex spectrum. Similarly we have shown that all PT-symmetric $G_s$-pseudo Hermitian matrices must have real eigenvalues. In section~\ref{sec:4}, from the definitions of $G$- and $G_s$-pseudo Hermitian matrices we have constructed seven ensembles of $2\times2$ $G(G_s)$-pseudo Hermitian matrices (or equivalently ensembles of PT-symmetric matrices that are having four real parameters), which represent all possible $2\times2$ $G(G_s)$ pseudo Hermitian matrices. In other words for a given $G$ ($G_s$), the set of all  $2\times2$  $G$ ($G_s$)-Pseudo Hermitian can be thought of in terms of a linear variety. For a singular $G_s$ the collection of all $H$ satisfying
$H^\dagger G_s = G_sH$ is a vector space of dimension of 5, such that a 4-dimensional subspace of PT-symmetric matrices exists. 
It is further found that for any two Hermitian $G_i, G_j\in M_2(\mathbb{C})$ such that $G_i\neq \lambda G_j$, there always exists exactly one trace less PT-symmetric $H\in M_2(\mathbb{C})$ (up to real scaling ) which is pseudo-Hermitian 
with respect to both these $G$ matrices. 

The determinants of each one of the seven ensembles of trace-less G-pseudo Hermitian matrices, denoted by $H_i$ ($i=1,2,...,7$), are studied in section~\ref{sec:5}. In all seven cases, when $\det{G}<0  $, the cases $\det{H_i} = \lambda$, 
$\det{H_i} = 0$ and $\det{H_i} = -\lambda$ : $\lambda>0$, 
respectively define  hyperboloid of 1 sheet, quadric cone, and hyperboloid of 2 sheets. All the points on hyperboloid of one sheet represent PT-symmetric matrices with complex eigenvalues (broken PT-symmetry), and all the points on  quadric cone and hyperboloid of 2 sheets  represent PT-symmetric matrices with real eigenvalues (unbroken PT-symmetry). On the other hand, when $\det{G} >0$, the $\det{H_i} = -\lambda $, defines an ellipsoid. In this case all the points on the ellipsoid represent PT-symmetric matrices with real eigenvalues (unbroken PT-symmetry). Similarly, in section~\ref{sec:6}, we have shown that for singular $G_s$, quadrics associated with the determinant of ensembles of trace-less  PT-symmetric matrices (denoted by $H'_i$ ($i=1,2,...,7$),) define two parallel planes, when $\det{H'_i} = -\lambda$ and a single plane when 
$\det{H'_i} = 0$. All the points on two parallel planes, and  the single plane  represent PT-symmetric matrices with real eigenvalues (unbroken PT-symmetry). Set of all Hermitian matrices $G\in M_2(\mathbb{C})$, satisfying $H^{\dagger}G = GH$, given a specific PT-symmetric $H\in M_2(\mathbb{C})$, turns out to be describable in terms of quadratic varieties. In section~\ref{sec:7}, we have shown that all $G \in M_2(\mathbb{C})$ with a fixed trace for a given $H \in M_2(\mathbb{C})$, satisfying  $H^{\dagger}G = GH$ can be described in terms of the intersection of six quadrics. The possible quadric structures are found to be among the quadrics: quadric cone, hyperboloid of 1 sheet, hyperboloid of 2 sheets, hyperbolic paraboloid, cylinder, hyperbolic cylinder, parabolic cylinder, two planes passing through the origin, two parallel planes, a single plane, and a straight line. This brings out some geometrical  structures associated with such ensembles of PT-symmetric matrices thereby possibly leading to better insights into the behavior of 
the systems that they represent.

We begin by stating some general definitions and notations.

Any $H\in M_2(\mathbb{C})$ can be written as a linear combination of Pauli matrices $\sigma_1,\sigma_2,\sigma_3$ and the identity matrix $\sigma_0=I$ as 

\begin{equation}\label{Eq.(1.1)}
H = \sigma_{0}\,h^{0} + \pmb{\sigma \cdot h} = \left[\begin{array}{cc}
h^0 + h^3  & h^1 - ih^2 \\
h^1 + ih^2 & h^0 - h^3
\end{array}\right]
\end{equation} 
Here,
$h^0=h^0_R+ih^0_I, \,\pmb{\sigma}=(\sigma_1,\sigma_2,\sigma_3), \,
\mathrm{and} \,\,
\pmb{h}=\pmb{h}_R+i\pmb{h}_I =(h^1_R + ih^1_I, h^2_R + ih^2_I, h^3_R + ih^3_I)$, such that \,$\pmb{h}\in\mathbb{C}^3$ and \,$\pmb{h}_R,\pmb{h}_I\in\mathbb{R}^3$ . 
The characteristic polynomial of $H$ and its roots, respectively, are given by
\begin{eqnarray}
f(E) &=& E^2-2h^0E+(h^0)^2-\pmb{h}\cdot\pmb{h}
\\
\label{eq:E}
 E&=&h^0\mp\sqrt{\pmb{h}\cdot\pmb{h}}
\end{eqnarray}
where $\pmb{h}\cdot\pmb{h}=\pmb{h}_R\cdot\pmb{h}_R-\pmb{h}_I\cdot\pmb{h}_I+2i \pmb{h}_R\cdot\pmb{h}_I$

If Hamiltonian $H$ is PT-Symmetric, then $\left[H, PT\right] = 0$, where $P$ is a parity operator and T is a time reversal operator. For the two-level systems, $\left[H, PT\right]=0$ if and only if $H$ has a real characteristic polynomial \cite{Bender:2009mq}. Then, from Eq.~(\ref{Eq.(1.1)}), it follows that $H$ is PT-symmetric if and only if  $h^0_I = 0$ and $\pmb{h}_R\cdot\pmb{h}_I = 0$
\cite{Bender:2009mq}. Thus any $2\times 2$ PT-symmetric matrix can be characterized by six real parameters. 
We denote the set of all $2\times 2$  PT-symmetric matrices by $S_{PT}$.

\section{A Partition of $S_{PT}$}\label{sec:2}
Keeping $h^0_I = 0$ and $h^0_R\in\mathbb{R}$, we consider four distinct cases in which $\pmb{h}_R \cdot \pmb{h} _I = 0$. This naturally gives rise to a partition of the set $S_{PT} = S_{1} \cup S_{2} \cup S_{3} 
\cup S_{4}$ such that $S_{i} \cap S_{j} = \emptyset$ for 
$i \ne j$. 
\subsection{Case-1: $\pmb{h}_R\neq 0 $ and   $\pmb{h}_I= 0$} 
From Eq.~(\ref{Eq.(1.1)}), it follows that $H^\dag=H\iff h^0_I=0$ and $\pmb{h}_I=0$. Thus all matrices belonging to this set (denoted by $S_1$) are Hermitian matrices with non-degenerate eigenvalues. All these 
matrices represent unbroken PT-symmetric operators. Any $H \in S_1$  can be written as
\begin{equation}
H = \sigma_{0}\,h_R^{0}+\pmb{\sigma \cdot h}_{R} = \left[\begin{array}{cc}
h^0_R + h^3_R  & h^1_R - ih^2_R \\
h^1_R + ih^2_R & h^0_R - h^3_R
\end{array}\right]
\end{equation}
\subsection{Case-2 $\pmb{h}_R = 0$
and $\pmb{h}_I\neq 0$} 
All matrices belonging to this set (denoted by $S_2$)  represent broken PT-symmetric operators and are $SU(2)$-like matrices with positive determinant and complex conjugate pairs of eigenvalues. From Eq.~(\ref{Eq.(1.1)}), it follows that $H^\dag=-H \iff h^0_R = 0$ and $\pmb{h}_R = 0$. The trace-less matrices in this set form the set of all $2\times 2 $ non-zero PT-symmetric anti-Hermitian matrices and they together with zero matrix form the Lie algebra of $SU(2)$. 
Any $H\in S_2$ can be written as
\begin{equation}\label{eq:Hs2}
H = \sigma_{0}\,h_R^{0} + i\,\pmb{\sigma \cdot h}_{I} = \left[\begin{array}{cc}
h^0_R + ih^3_I  & h^2_I +ih^1_I \\
-h^2_I +ih^1_I &  h^0_R -ih^3_I
\end{array}\right] = \left[\begin{array}{cc}
\alpha  & -\beta^* \\
\beta & \alpha^*
\end{array}\right]
\end{equation}
     
\subsection{Case-3: $\pmb{h}_R = 0$
and $\pmb{h}_I = 0$}
All matrices belonging to this set (denoted by $S_3$)  represent unbroken PT-symmetric operators with degenerate eigenvalues. Any $H\in$ $S_3$ can be written as
\begin{equation}
H= \sigma_{0}\,h_R^{0}
\end{equation}
with $\pmb{h} \cdot \pmb{h} = 0$ $\forall H \in S_3$.

From the discussions above, it is clear that 
the set $S_1\cup S_3$ gives the set of all $H\in M_2(\mathbb{C})$ such that $H^\dag = H$. This set has a 4-dimensional vector space structure over $\mathbb{R}$.

The set  $S_2\cup S_3$ is the set of all $H\in M_2(\mathbb{C})$ such that $H^\dag H = H H^\dag = \sigma_0 \det(H)$. This set has a 4-dimensional vector space structure over $\mathbb{R}$. This statement will be proved 
below.

The set $S_2\cup (S_3-0\sigma_0)$ is the set of all $2\times2$ $SU(2)$-like matrices and is a group under multiplication where $SU(2)$ is its normal subgroup. The proof of this  statement
follows from definition of $SU(2)$-like matrices.

The set $S_1\cup S_2\cup S_3$ is the set of all  $2\times 2 $ normal PT-symmetric matrices satisfying $\pmb{h}_R \times \pmb{h} _I = 0$. In this connection, the following result is of interest:
 \begin{prop}\label{prop:normal}
Any $H\in M_2(\mathbb{C})$ is normal if and 
only if $\pmb{h}_R \times  \pmb{h}_I = 0$.
\end{prop}
\begin{proof}
Any $H\in M_2(\mathbb{C})$ is normal if and only if  $\left[H,H^{\dag}\right] = 0$ i.e. 
$ HH^{\dag} = H^{\dag}H $.\\
Let 
\begin{equation}\label{eq:hhdagger}
HH^{\dag} = \mathbb{H}= \sigma_{0}\,\mathfrak{h}^{0} + \pmb{\sigma \cdot \mathfrak{h}}
\end{equation}
 and 
 \begin{equation}\label{eq:hdaggerh}
H^{\dag}H = \tilde{\mathbb{H}} = \sigma_{0}\,\mathfrak{\tilde{h}}^{0} + \pmb{\sigma \cdot \mathfrak{\tilde{h}}} 
 \end{equation}
 Now by making use of Eq.~(\ref{Eq.(1.1)}) in Eq.~(\ref{eq:hhdagger}) and Eq.(\ref{eq:hdaggerh}), we get the following equations.
 \begin{eqnarray}
 \label{eq:2.6}
 \mathfrak{h}^0_R & = &\mathfrak{\tilde{h}}^0_R =(h^0_R)^2 + (h^0_I)^2 + |\pmb{h}_R|^2 + |\pmb{h}_I|^2
\\
\mathfrak{h}^0_I & = &\mathfrak{\tilde{h}}^0_I = 0
\\ 
\label{eq:2.8}
\pmb{\mathfrak{h}}_R & = & 2\left[h^0_R \pmb{h}_R + h^0_I \pmb{h}_I + \pmb{h}_R\times \pmb{h}_I\right]
\\
\pmb{\mathfrak{\tilde{h}}}_R & = & 2\left[h^0_R \pmb{h}_R + h^0_I \pmb{h}_I - \pmb{h}_R\times \pmb{h}_I\right]
\\
\label{eq:2.10}
\pmb{\mathfrak{h}}_I & = & \pmb{\mathfrak{\tilde{h}}}_I = 0 
\end{eqnarray}
By putting the Eq.~(\ref{eq:2.6}) - Eq.~(\ref{eq:2.10}) in the condition $\mathbb{H}=\mathbb{\tilde{H}}$ the required proof is obtained.
\end{proof}
In order to establish the second statement, we observe that:
 \begin{prop}
 Let $H\in M_2(\mathbb{C})$. Then $H$ satisfies
 $H^\dag H = H H^\dag = \sigma_0\det(H)$
if and only if  $H \in S_2\cup S_3$.
\end{prop}
\begin{proof}
From proposition~\ref{prop:normal}, we know that  $H^\dag H =H H^\dag  \iff \mathbf{h}_R \times  \mathbf{h}_I = 0$.
\\Then $\mathbb{H} = \mathbb{\tilde{H}} = \sigma_{0}\,\mathfrak{h}^{0}_R + \pmb{\sigma \cdot \mathfrak{h}}_R$, where $\mathfrak{h}^0_R$ and $\pmb{\mathfrak{h}}_R$ given by Eq.~(\ref{eq:2.6}) and Eq.~(\ref{eq:2.8}) with $\mathbf{h}_R \times  \mathbf{h}_I = 0$.
Now 
\begin{equation}
 \sigma_0\det(H) = \mathbb{H} = \mathbb{\tilde{H}} 
\iff h^0_I = 0  \ and \
\mathbf{h}_R = 0
\end{equation}
 \end{proof}
In addition, 
\begin{prop} 
For any $U$, $V$ belonging to $S_2\cup S_3$, the following
relations hold:
 \begin{eqnarray}\label{eq:uv}
  U V^\dag + V U^\dag = U^\dag V + V^\dag U = \sigma_0\, 
 \frac{1}{2}\, \mathrm{Tr}(U)\, \mathrm{Tr}(V) + 2\, \sigma_0 (\mathbf{u}_I\cdot\mathbf{v}_I) \\
 \label{eq:u+v}
 \det(U + V) = \det(U) + \det(V) + \frac{1}{2} \mathrm{Tr}(U)\, \mathrm{Tr}(V) + 2\, (\mathbf{u}_I \cdot \mathbf{v}_I) \\
 \label{eq:u+vdagger}
(U + V)(U + V)^\dag = (U + V)^\dag\, (U + V) = \sigma_0 \det(U + V)
\end{eqnarray}
\end{prop}
\begin{proof}
\begin{enumerate}
\item From Eq.~(\ref{eq:Hs2}), we have $U= \sigma_{0}\,u_R^{0} + i\,\pmb{\sigma \cdot u}_{I}$ and 
$V= \sigma_{0}\,v_R^{0} + i\,\pmb{\sigma \cdot v}_{I}$.Let $UV^\dag = W$ and $U^\dag V = \tilde{W}$, then
\begin{equation}
W = \sigma_{0}\,w_R^{0} + i\,\pmb{\sigma \cdot w}_{I}
\end{equation}
where 
\begin{equation}
w_R^{0} = u^0_R\, v^0_R + \mathbf{u}_I\cdot\mathbf{v}_I
\end{equation} and
\begin{equation}
\mathbf{w}_I = v^0_R\,\mathbf{u}_I-u^0_R\,\mathbf{v}_I + \mathbf{u}_I\times\mathbf{v}_I
\end{equation}
From the above equations we get 
\begin{equation}\label{eq:w+wdagger}
    W + W^\dag = 2 \sigma_0 \,w^0_R
\end{equation} 
Similarly, we can show that
\begin{equation}\label{eq:wtilde+wrildedagger}
  \tilde{W} + \tilde{W}^\dag = 2 \sigma_0\,w^0_R = \sigma_0\, 
 \frac{1}{2} \,\mathrm{Tr}(U)\, \mathrm{Tr}(V) + 2\, \sigma_0 (\mathbf{u}_I\cdot\mathbf{v}_I)
\end{equation} 
For any $U$, $V$ belonging to $S_3$, all the above equations will hold with $\mathbf{u}_I = \mathbf{v}_I = 0$, hence from Eq.~(\ref{eq:w+wdagger}) and Eq.~\ref{eq:wtilde+wrildedagger} the proof of Eq.~(\ref{eq:uv}) is obtained.
\item Let $X=U+V$ and $X$ can be expressed as Eq.~(\ref{Eq.(1.1)}) with corresponding coefficients of $\sigma$'s obtained by component-wise addition of coefficients of U and V. Then it is straightforward to show 
\begin{equation}\label{eq:detX}
 \det{X}=(x^0)^2-\mathbf{x}\cdot\mathbf{x}= \det(U) + \det(V) + \frac{1}{2} \mathrm{Tr}(U)\, \mathrm{Tr}(V) + 2 (\mathbf{u}_I \cdot \mathbf{v}_I) 
\end{equation}
The proof of Eq.~(\ref{eq:u+v}) is established.
\item \begin{equation}
    XX^\dag=X^\dag X=\sigma_0[\det{U}+\det{V}]+W+W^\dag
\end{equation}
From Eq.~(\ref{eq:w+wdagger}) and Eq.~(\ref{eq:wtilde+wrildedagger}), it follows that 
\begin{equation}\label{eq:XXdagger}
 XX^\dag=\sigma_0[ \det(U) + \det(V) + \frac{1}{2} \mathrm{Tr}(U) \mathrm{Tr}(V) + 2 (\mathbf{u}_I \cdot \mathbf{v}_I) ]
\end{equation}
Eq.~(\ref{eq:detX}) in Eq.~(\ref{eq:XXdagger}) gives
\begin{equation}
    XX^\dag=X^\dag X=\sigma_0 \det{X}
\end{equation}
\end{enumerate}
Eq.~(\ref{eq:u+vdagger}) is proved
\end{proof}
The second result stated at the beginning of the section
follows directly from Eq.~(\ref{eq:u+vdagger}).

\subsection{Case-4: $\pmb{h}_R\neq 0 $ and   $\pmb{h}_I\neq 0$ with $\pmb{h}_R\cdot\pmb{h}_I=0$}

All matrices belonging to this set (denoted by $S_4$) are non-normal PT-symmetric matrices. Matrices in the set can have real and distinct eigenvalues, complex conjugate eigenvalues, and real and degenerate eigenvalues  depending on the following conditions respectively 1) $|\pmb{h}_R|>|\pmb{h}_I|$,
2) $|\pmb{h}_R|<|\pmb{h}_I|$ and
3) $|\pmb{h}_R|=|\pmb{h}_I|$.

In the case of the condition  $|\pmb{h}_R|>|\pmb{h}_I|$, the corresponding matrices represent unbroken PT-symmetric operators with non-degenerate eigenvalues and they are all diagonalizable. On the other hand for $|\pmb{h}_R|<|\pmb{h}_I|$, the matrices represent broken PT-symmetric operators and they have complex conjugate eigenvalues and are also diagonalizable. Finally, the
matrices with $|\pmb{h}_R|=|\pmb{h}_I|$  represent unbroken PT-symmetric operators with degenerate eigenvalues and are non-diagonalizable. In fact, these matrices are the only PT-symmetric matrices having Jordan block matrix form. Notice that the condition $\pmb{h}\cdot\pmb{h}=0$ is satisfied for all such matrices having Jordan form. The classification thus obtained allows us to systematically
study the set $S_{PT}$.
The results are summarised in Table~\ref{tab:1}.

\begin{table}[htb]
\centering
\caption{Properties of matrices belonging to the classes
$S_1$, $S_2$, $S_3$ and $S_4$.}
\label{tab:1}
\begin{tabular}{|c|c|c|c|}\hline
 Set & Nature of Eigenvalues     & Normal/Non-normal &Diagonalizability \\\hline
 $S_1$  & Real, non-degenerate & Normal & Diagonalizable \\\hline
$S_2$  & Complex conjugate    & Normal & Diagonalizable \\\hline
$S_3$  & Real, degenerate     & Normal & Diagonalizable \\\hline
$S_4$  & Real, non-degenerate    & Non-normal &Diagonalizable \\\hline
 $S_4$  & Complex conjugate   & Non-normal &Diagonalizable \\\hline
$S_4$  & Real, degenerate    & Non-normal &Non-diagonalizable \\\hline
\end{tabular}
\vskip\baselineskip
\end{table}

\section{$G(G_s)$-Pseudo-Hermitian Matrices}\label{sec:3}

 In this section we consider $H\in M_2(\mathbb{C})$ satisfying  the equation,  \begin{equation}\label{eq:HdaggerG}
    H^{\dag}G = G H 
 \end{equation} with  Hermitian $G \in M_2(\mathbb{C})$. For any $H$ satisfying  Eq.~\ref{eq:HdaggerG} with a given $G$,  we have the following equations.
 \begin{equation}\label{eq:g0Rh0I}
g^0_{R} \pmb{h}_{I} + h^0_{I} \pmb{g}_{R} + \pmb{g}_R \times \pmb{h}_R = 0 \quad\text{and}\quad g^0_R h^0_I= - \pmb{h}_I\cdot\pmb{g}_R  
\end{equation}
 
An equivalent system of equations can be obtained from Eq.~(\ref{eq:g0Rh0I}) and is given by 
\begin{equation}\label{eq:hRgR2}
\pmb{h}_R\left|\pmb{g}_R\right|^2 = \pmb{g}_R\left(\pmb{h}_R\cdot\pmb{g}_R\right) + g^{0}_R \left(\pmb{g}_R \times \pmb{h}_I\right)
\end{equation} 
and\begin{equation}\label{eq:hIg0R2}
\pmb{h}_I\left(g^{0}_R\right)^2= \pmb{g}_R\left(\pmb{h}_I\cdot \pmb{g}_R\right) + g^{0}_R\left(\pmb{h}_R \times \pmb{g}_R\right) 
\end{equation}
When $H$ satisfies Eq.~(\ref{eq:HdaggerG})  with a $G_s$   such that $\left|g^{0}_R\right| = \left|\pmb{g}_R\right| \neq0$, then $h^0_I$ is not equal to zero in general. This clearly indicates that all $H$ satisfying Eq.~(\ref{eq:HdaggerG}) with a singular $G_s$ are not PT-symmetric.

A matrix $H\in M_2(\mathbb{C})$ is called $G$-Pseudo Hermitian if there exists a non-singular Hermitian matrix $G$ such that Eq.~(\ref{eq:HdaggerG}) is satisfied, i.e,
\begin{equation}\label{eq:Hdagger}
H^\dag = GHG^{-1}
\end{equation}
where, $\left|\pmb{g}_R\right| \neq \left|g^0_R\right|$. The Eq.~(\ref{eq:Hdagger}) is satisfied if and only if  $ h^{0}_I = 0$ and 
$G(\pmb{\sigma \cdot h})G^{-1} = (\pmb{\sigma \cdot h})^\dag$. 
We have the following propositions for $G$- pseudo hermitian matrices.

\begin{prop}
Any matrix $H\in M_{2}(\mathbb{C})$ is $G$-pseudo-Hermitian only if it is PT-symmetric.
\end{prop}
\begin{proof}
   From Eq.~(\ref{eq:hIg0R2}) we get 
\begin{equation}\label{eq:hI.gR*g0R2}
\left(\pmb{h}_I\cdot\pmb{g}_R\right)\left(g^{0}_R\right)^2= \left|\pmb{g}_R\right|^2\left(\pmb{h}_I\cdot\pmb{g}_R\right)
\end{equation}
Since $\left|\pmb{g}_R\right| \neq \left|g^0_R\right|$ , Eq.~(\ref{eq:hI.gR*g0R2}) is valid only if
\begin{equation}\label{eq:hI.gR=0}
\pmb{h}_I\cdot\pmb{g}_R = 0
 \end{equation} 
From Eq.~(\ref{eq:hRgR2}), Eq.~(\ref{eq:hIg0R2}), and Eq.~(\ref{eq:hI.gR=0}) we get 
\begin{equation}\label{eq:hI.hR=0}
\pmb{h}_I\cdot\pmb{h}_R = 0
\end{equation}  
the  Eq.~(\ref{eq:hI.hR=0}) and $h^0_I=0$, are  precisely the conditions for PT-symmetric matrices.
\end{proof}
It has been proved \cite{Zhang:2019gyc} that all PT-symmetric matrices belonging 
to $M_{n}(\mathbb{C})$ are pseudo-Hermitian. Combining this result
with the result above, we arrive at the equivalence of PT-symmetric 
and pseudo-Hermitian matrices for $M_{2}(\mathbb{C})$. Precisely:
\begin{thm}
The matrix $H\in M_{2}(\mathbb{C})$ is $G$-pseudo-Hermitian if and only if it is PT-symmetric.
\end{thm}

From the above theorem it is clear that for every $2\times2$ PT-symmetric matrix there exists a $G$ such that Eq.~(\ref{eq:Hdagger}) is satisfied. 

\begin{prop}\label{prop:HS1}
 Any $H\in S_1$ is  $G$-psuedo-Hermitian  only if $\pmb{h}_R \times \pmb{g}_R =0$
\end{prop}
\begin{proof}
From Eq.~(\ref{eq:hRgR2}), for any $H\in S_1$  we get 
\begin{equation}
 \pmb{h}_R \times \pmb{g}_R =0 \implies  \left[H,G\right] = 0
\end{equation}
\end{proof}

\begin{prop}\label{prop:HS2}
  Any $H\in S_2$ is  $G$-pseudo Hermitian only if $g^0_R =0$ (i.e $G$ is trace less).
\end{prop}
\begin{proof}
 For any $H\in S_2$, Eq.~(\ref{eq:hRgR2}) yields $g^{0}_R \left(\pmb{g}_R \times \mathbf{h}_I\right)=0$. Then it follows 
 from invertibility of $G$ and Eq.~(\ref{eq:hI.gR=0}) that either $g^0_R =0$ or  $\pmb{g}_R=0$. If $\pmb{g}_R=0$,  Eq.~(\ref{eq:hIg0R2}) $\implies g^{0}_R = 0$, which is not possible since $G$ is invertible. Then the only  possibility is  $g^{0}_R=0$ and $\pmb{g}_R\neq0$, which proves the result.
\end{proof}

 \begin{prop}\label{prop:HS4}
For any $G$-psuedo-Hernitian  $H\in S_4$,
\begin{equation}
\pmb{h}_R \times \pmb{g}_R=0 \iff g^0_R =0
\end{equation} 
\end{prop}
\begin{proof}
For any  $H\in S_4$, and when $Trace(G) = 0$, from Eq.~(\ref{eq:hRgR2}) we can say that  $\pmb{h}_R\cdot \pmb{g}_R$ must be non-zero, further from the same equation we can see that $\pmb{h}_R \times \pmb{g}_R$ must be zero. Similarly when $\pmb{h}_R \times \pmb{g}_R=0$, Eq.~(\ref{eq:hIg0R2}) and Eq.~(\ref{eq:hI.gR=0}) imply  that $Trace(G) = 0$.
\end{proof}
\begin{prop}
For any $G$-psuedo-Hermitian $H\in S_4$,  $\pmb{h}_R\cdot\pmb{g}_R=0$, i.e., the  vectors $\pmb{h}_R,\pmb{h}_I,\pmb{g}_R$ are all mutually perpendicular only if $ Trace(G)\neq0 $  and H is diagonalizable.
\end{prop}

\begin{proof}
When  $\pmb{h}_R\cdot\pmb{g}_R=0$, Eq.~(\ref{eq:hRgR2}) $\implies$
$\pmb{h}_R\left|\pmb{g}_R\right|^2= g^{0}_R \left(\pmb{g}_R \times \pmb{h}_I\right)$, taking the magnitude on both sides,  we get 
\begin{equation}
\frac{\left|\pmb{h}_R\right|}{\left|\pmb{h}_I\right|} = \frac{\left|g^0_R\right|}{\left|\pmb{g}_R\right|}\neq1
\end{equation}
\end{proof}
Thus for any $G$-pseudo Hermitian $H\in S_4$  can have a  Jordan block form only if $\pmb{h}_R\cdot\pmb{g}_R\neq0$.
\begin{prop}
For any $G$-psuedo-Hermitian $H\in S_4$, such that $Trace(G)\neq0$ and  $\pmb{h}_R\cdot\pmb{g}_R\neq0$, then the matrix H is non-diagonalizable if and  only if, the angle $\theta$ between vectors  $\pmb{h}_R $ and $\pmb{g}_R $ is
\begin{equation}
 \theta = \sin^{-1}\left( \frac{\left|g^0_R\right|}{\left|\pmb{g}_R\right|}\right)
\end{equation}
\end{prop}
\begin{proof}
In general when H$\in S_4$ and $g^0_R\neq0$, from Eq.~(\ref{eq:hIg0R2}) we get 
 \begin{equation}\label{eq:modsintheta}
\left|\sin(\theta)\right|=\frac{\left|\pmb{h}_I\right|\left|g^0_R\right|}{\left|\pmb{h}_R\right|\left|\pmb{g}_R\right|}
    \end{equation}
$\implies$
\begin{equation}
0<\frac{\left|\pmb{h}_I\right|\left|g^0_R\right|}{\left|\pmb{h}_R\right|\left|\pmb{g}_R\right|}\leq1
\end{equation}
Equality holds only when $\theta = \pm \pi/2$.
Any $H\in S_4$ is non-diaganolizable if and only if 
$ \left|\pmb{h}_R\right| = \left|\pmb{h}_I\right|  $. This condition in Eq.~(\ref{eq:modsintheta}) establishes the proof.
\end{proof}
\begin{prop}
    Any $G$-psuedo-Hermitian $H\in M_{2}(\mathbb{C})$, such that $Trace(G)\neq0$  and $\det(G)>0$, has real eigenvalues  
\end{prop}
\begin{proof}
   From Eq.~(\ref{eq:g0Rh0I}), when $h^0_I = 0$, we have 
   \begin{equation}
     g^0_{R} \pmb{h}_{I} + \pmb{g}_R \times \pmb{h}_R = 0   
   \end{equation}
 $\implies$
 \begin{equation}\label{eq:modhRbymodhI}
     \frac{\left| \pmb{h}_R\right|}{\left| \pmb{h}_I\right|} = \frac{\left|(g^0_{R})\right|}{\left|\pmb{g}_R\right| \left|\sin(\theta)\right|}
 \end{equation}

where $\theta $ is the angle between $\pmb{h}_R$ and $\pmb{g}_R$. If $\det(G) = (g^0_R)^2 - \left|\pmb{g}_R\right|^2 >0$, then 
 \begin{eqnarray}
    \left|\pmb{h}_R\right| >\left|\pmb{h}_I\right|  
 \end{eqnarray}.
 
\end{proof}
\begin{prop}
  Any $G$-psuedo-Hermitian $H\in M_{2}(\mathbb{C})$, such that $Trace(G)\neq0$  and $\det(G)<0$, can have real or complex conjugate pairs of  eigenvalues.  
\end{prop}
\begin{proof}
    From Eq.~(\ref{eq:modhRbymodhI}), when $\det(G) < 0$ , then we have all the possibilities $  \left|\pmb{h}_R\right| >\left|\pmb{h}_I\right|,  \left|\pmb{h}_R\right| = \left|\pmb{h}_I\right|, \left|\pmb{h}_R\right| <\left|\pmb{h}_I\right|$, corresponding to real and distinct , real and degenerate and complex conjugate pairs of eigenvalues. 
 \end{proof}
 In section~\ref{sec:5} we will show that proposition 3.9 is true in general even when $Trace(G)=0$

We have the following propositions for $G_s$- pseudo hermitian matrices.
 \begin{prop}
Any $H$ satisfying the equation $H^{\dag}G_s = G_s H $ is PT-symmetric if and only if $\pmb{h}_I\cdot\pmb{g}_R=0$
\begin{proof}
Eq.~(\ref{eq:g0Rh0I}), gives 
\begin{equation}\label{eq:hI.hR}
\pmb{h}_I\cdot\pmb{h}_R = \frac{(\pmb{h}_I\cdot\pmb{g}_R) (\pmb{h}_R\cdot\pmb{g}_R) }{\left|\pmb{g}_R\right|^2} 
\end{equation}
from the second condition of  Eq.~(\ref{eq:g0Rh0I}) and from Eq.~(\ref{eq:hI.hR}), $H$ is PT-symmetric if and only if 
$\pmb{h}_I\cdot\pmb{g}_R =0 $.
\end{proof}
\end{prop} 

Note that for any $H\in S_1$ to satisfying Eq.~(\ref{eq:HdaggerG}) with a singular $G_s$, the condition $ \left[H,G_s\right] = 0$ holds as in Proposition 3.3.
\begin{prop}
There exists no $H \in S_2$ satisfying the equation $H^{\dag}G_s = G_sH$ with a singular Hermitian $G_s \neq0$.
\begin{proof}
When $H\in S_2$,  Eq.~(\ref{eq:g0Rh0I}) implies $g^0_R =0$, which contradict the assumption that $g^0_R \neq0$.
\end{proof}
\end{prop}
 \begin{prop}
Any $H\in S_4$ satisfying $H^{\dag}G_s = G_s H$ with a singular Hermitian $G_s$ has real eigenvalues.
\begin{proof}
When $H\in S_4$, from Eq.~(\ref{eq:g0Rh0I})  we have 
\begin{equation}\label{eq:modsintheta2}
\left|\sin(\theta)\right|=\frac{\left|\pmb{h}_I\right|}{\left|\pmb{h}_R\right|} \implies \left|\pmb{h}_R\right| \geq \left|\pmb{h}_I\right|
\end{equation}
where $\theta$ is the angle between vectors  $\pmb{h}_R $ and $\pmb{g}_R $, it follows from Eq.~(\ref{eq:modsintheta2}) and Eq.~(\ref{eq:E}), that $H$ has real eigenvalues. When  $\theta = \frac{\pi}{2}$, $H$ becomes non diagonalizable.
\end{proof}
\end{prop}
From all the propositions above we have the following result. 

\begin{thm}
Every $2 \times 2$ PT-symmetric matrix $H$ satisfying the equation $H^{\dag}G= GH $ with a singular Hermitian $G$ has real spectrum or equivalently has unbroken PT-symmetry.
\end{thm}

\begin{thm}
 For any  two Hermitian $G_i,G_j\in M_2(\mathbb{C})$ such that $G_i\neq \lambda G_j$, $\lambda \in \mathbb{R}$ there always exists exactly one trace less PT-symmetric $H\in M_2(\mathbb{C})$ (up to scaling ) which is pseudo-Hermitian with respect to both these $G$ matrices.  
\end{thm}
 \begin{proof}
 Suppose that $H\in M_2(\mathbb{C})$ satisfies Eq.~(\ref{eq:HdaggerG}) with two Hermitian matrices G and F such that $g^0_R \neq0\neq f^0_R$ . Then from Eq.~(\ref{eq:g0Rh0I}) we get
  \begin{equation}\label{eq:h0Iw}
  h^0_I\,\pmb{w} = \pmb{h}_R\times\pmb{w}
  \end{equation} where 
  \begin{equation}
      \pmb{w} =  f^0_{R}\, \pmb{g}_R -  g^0_R\,\pmb{f}_R
  \end{equation}
  when $h^0_I\neq0$ (i.e, $\det(G) = 0 = \det(F) $, the Eq.~(\ref{eq:h0Iw}) is true only when \pmb{w} = 0, which implies F and G must be the same upto scaling. Thus when $h^0_I\neq0$, the non-PT-symmetric $H$ must satisfy Eq.~(\ref{eq:HdaggerG}) with a unique singular $G$ (upto scaling). When $h^0_I = 0$,  from Eq.~(\ref{eq:h0Iw}), we get 
  \begin{equation}\label{eq:lmabdaw}
  \pmb{h}_R = \lambda \,\pmb{w} ;\, \lambda \in \mathbb{R}
  \end{equation}
 Similarly when $h^0_I = 0$, from Eq.~(\ref{eq:g0Rh0I}) we get 
  \begin{equation}\label{eq:hRcrossgR}
\pmb{h}_I\,g^0_{R} = \pmb{h}_R\times\pmb{g}_R 
  \end{equation}
  and
  \begin{equation}\label{eq:hRcrossfR}
        \pmb{h}_I\,f^0_{R} =      \pmb{h}_R\times\pmb{f}_R  
  \end{equation}
  making use Eq.~(\ref{eq:lmabdaw}) in Eq.~(\ref{eq:hRcrossgR}), and Eq.~(\ref{eq:hRcrossfR}), we get 
  \begin{equation}\label{eq:gRcrossfR}
      \pmb{h}_I = \lambda\quad \pmb{g}_R \times \pmb{f}_R
  \end{equation}
    That is when $H$ is trace-less and if it satisfies Eq.~(\ref{eq:HdaggerG}) with $G$ and $F$, then $\pmb{h}_R $  and $\pmb{h}_I$ are given by Eq.~(\ref{eq:lmabdaw}) and Eq.~(\ref{eq:gRcrossfR}).
  For any two Hermitian matrices $G$ and $F$ with associated vectors $\pmb{g}_R ,\pmb{f}_R \in\mathbb{R}^3 $, there always exists a $\pmb{h}_R ,\pmb{h}_I \in\mathbb{R}^3$ given by Eq.~(\ref{eq:lmabdaw}) and Eq.~(\ref{eq:gRcrossfR}) , which can be associated with a trace less PT-symmetric $ H\in M_{2}(\mathbb{C})$. This result is generally true even when $g^0_R =0= f^0_R$.
  \end{proof}

  \begin{thm}
   For any  two invertible Hermitian matrices $G,F\in M_2(\mathbb{C})$ such that $G\neq \lambda F$, $\lambda \in \mathbb{R}$ there always exists exactly one trace less PT-symmetric  $H\in M_2(\mathbb{C})$ (up to scaling by a constant) which is pseudo-Hermitian with respect to both these $G$ matrices such that $\left[ H,G^{-1}F\right ]= 0 = \left[ H,F^{-1} G \right]$.  
  \end {thm}
  \begin{proof}
      The proof is straight forward from theorem 3.14 and the definition of G-Pseudo Hermitian matrices.
  \end{proof}

\section{Ensembles of G($G_s$)-Pseudo Hermitian Matrices}\label{sec:4}

In this section, we explicitly construct  seven ensembles of all admissible matrices $H\in M_{2}\left(\mathbb{C}\right)$, corresponding to seven
 ensembles of invertible (singular) $G$ ($G_s$)  such that Eq.~(\ref{eq:HdaggerG}) is satisfied. These seven ensembles of matrices represent all possible $2\times2$ $G(G_s)$-pseudo Hermitian matrices. Since the $G$ ($G_s$)-matrices are Hermitian matrices,
they belong to the set $S_1\sqcup  S_3$. The set $S_3$ 
contains identity matrix up to a scaling. It is therefore sufficient to consider
only the cases of $G$ ($G_s$)$\in S_1$. For the following calculations set $S_1$ is
partitioned into seven cells, denoted by  $S_{1,c} = G_c \sqcup G_{s,c} $ ($c$ = 1 to 7). We shall consider  
the ensemble of invertible (singular) matrices from each of these cells, denoted by $G_c$ ($G_{s,c}$).
 All $H$ satisfying Eq.~(\ref{eq:HdaggerG})  for each set  $G_c$($G_{s,c}$), will be 
called as $G_c$($G_{s,c}$) -pseudo Hermitian matrices.

From Eq.~(\ref{eq:hRgR2}) and Eq.~(\ref{eq:hIg0R2}) we get the following homogeneous system of linear equations:
\begin{equation}\label{eq:MX=0}
MX=0
\end{equation}
Where $X^{T}=\left(h^1_R,h^2_R,h^3_R,h^1_I,h^2_I,h^3_I\right)=(\pmb{h}_{R},\pmb{h}_{I})$ and $M$ is the $6\times6$ real coefficient matrix, having block matrix form as given below and it depends on 4 real numbers $g^{1}_R=a,g^{2}_R=b, g^{3}_R=c,g^{0}_R=d$.  
\begin{equation}
M=\left[\begin{array}{cc}
M_1  & M_2\\
M_3 & M_4
\end{array}\right]
\end{equation}
Where
{
\renewcommand{\arraystretch}{1}
\begin{equation}
M_1=\left[\begin{array}{ccc}
b^2 + c^2 & -a b & -a c\\
-a b & a^2 + c^2 & -b c\\
-a c & -b c & a^2 + b^2
\end{array}\right]
\end{equation}
\vskip 0.5\baselineskip

\begin{equation}
M_2=\left[\begin{array}{ccc}
0 & c d & -b d\\
-c d & 0 & a d\\
b d & -a d & 0
\end{array}\right] = -M_3
\end{equation}
\vskip 0.5\baselineskip
\begin{equation}
 M_4=\left[\begin{array}{ccc}
d^2 - a^2 & -a b & -a c\\
-a b & d^2 - b^2 & -b c\\
-a c & -b c & d^2-c^2
\end{array}\right]
\end{equation}}

Notice that each of these set $G_c$($G_{s,c}$) is characterised by
a vector $\pmb{g}_{R,c}$,  where $\left|d\right| \neq \left|\pmb{g}_{R,c}\right| \neq0$ and  $\left|d\right| = \left|\pmb{g}_{R,c}\right| \neq0$  for $G_c$ and $G_{s,c}$ respectively. All $H = \sigma_{0}\,h^{0} + \pmb{\sigma \cdot h} $, matrices
satisfying Eq.~(\ref{eq:HdaggerG}) for invertible (singular) $G(G_s)$ can be obtained by finding the solution space of $MX=0$, for each set $G_c$ ($G_{s,c}$).
For the sake of clarity, given a set $G_c$, we consider a finer partition consisting of two cells such that: a) $\mathrm{Tr}(G)\neq0$ matrices ($d\neq0$) and b) $\mathrm{Tr}(G) = 0$ matrices ($d=0)$. In this way for a given $G(G_s)$, the set of all $G(G_s)$-pseudo Hermitian can be expressed as a linear variety.

\subsection{ $G_1$-Pseudo Hermitian Matrices}

The set $G_1$ is characterized by the vector 
$\pmb{g}_{R,1}$ = $( a\neq0, b\neq0,  c\neq0)$ and $ d \neq \pm|\pmb{g}_{R,1}| $. 
 The matrices in this set can be expressed
as
\begin{equation}
 G_1= \left\lbrace \left[\begin{array}{cc}
d + c  & a - i b \\
a + i b & d - c
\end{array}\right] :a, b,c \in \mathbb{R}^{*} , d \in \mathbb{R}      \right\rbrace
\end{equation}

\noindent
{\it Case-1}: $d\neq0$. The general solution for Eq.~(\ref{eq:MX=0}) in  this case can be written as 
\begin{equation}
X_1 = \alpha_1\, h^3_{R} + \beta_1\, h^2_{I} + \gamma_1\, h^3_{I}
\end{equation}
where $ h^3_{R,1}, 
 h^2_{I,1}, h^3_{I,1}\in\mathbb{R}$ are the free variables, and $\alpha_1,\beta_1,\gamma_1\in\mathbb{R}^{6}$  are the particular solutions and they form a basis for the solution space. Let $\alpha_1=(\pmb{\alpha}_{R,1},\pmb{\alpha}_{I,1})$,
$\beta_1 = (\pmb{\beta}_{R,1},\pmb{\beta}_{I,1})$, 
and $\gamma_1=(\pmb{\gamma}_{R,1},\pmb{\gamma}_{I,1})$, leading to
$\pmb{h}_{R} = \pmb{\alpha}_{R,1}\, h^3_{R} + \pmb{\beta}_{R,1}\, h^2_{I} + \pmb{\gamma}_{R,1}\, h^3_{I} $ and $\pmb{h}_{I,1} = \pmb{\alpha}_{I,1}\, h^3_{R} + \pmb{\beta}_{I,1}\, h^2_{I} +\pmb{\gamma}_{I,1}\, h^3_{I} $.
The particular solutions $\alpha_1$, $\beta_1$, and $\gamma_1$ are given as follows,

\begin{eqnarray}
\pmb{\alpha}_{1} = \left(\frac{a}{c}, \frac{b}{c},1,0,0,0\right);\quad\pmb{\beta}_{1} = \left(-\frac{d}{c}, -\frac{b\, d }{a \,c}, 0,-\frac{b}{a}, 1, 0\right);\quad\pmb{\gamma}_{1} = \left(0,-\frac{ d}{a}, 0,-\frac{c}{a}, 0, 1\right) 
\end{eqnarray}

Note that the vectors $\pmb{\alpha}_{R,1}$, $ \pmb{\alpha}_{I,1}$, $\pmb{\beta}_{I,1}$, $\pmb{\gamma}_{I,1}$ do not depend on $d$ and vectors  $\pmb{\beta}_{R,1}$, $\pmb{\gamma}_{R,1}$ depend  on $d$. The $2\times2$ matrix corresponding to $\alpha_1$ is PT-Symmetric normal and belongs to $S_1$, whereas  $2\times2$ matrices corresponding to $\beta_1$ and $\gamma_1$ are PT-Symmetric non-normal and belong to the set $S_4$. For a given $G \in G_c$ the particular solutions $\alpha,\beta,\gamma$ are fixed. The general ensemble of $G_1$-pseudo Hermitian matrix $H_1$ is given by
{
\renewcommand{\arraystretch}{1.3}
\begin{equation}
H_1=\left[\begin{array}{cc}
k_0 + k_1 + i k_3   &  k_1\,\frac{a}{c} + k_2 (1- \frac{d}{c}) - i \phi_1\\ k_1\,\frac{a}{c} -k_2(1 +\frac{d}{c}) + i \phi_2 &  k_0 - (k_1 + i k_3) \end{array}\right]
\end{equation}}

where 
\begin{align*}
\phi_1 &= k_1\,\frac{b}{c} + k_2\,\frac{b (1 - \frac{d}{c})}{a} + k_3\,\frac{(c - d)}{a},~ 
    \phi_2 = k_1\,\frac{b}{c} - k_2\,\frac{b (1 + \frac{d}{c})}{a} + k_3\,\frac{(c + d)}{a}
    \\
  k_0& = h^0_{R},~ k_1 = h^3_{R},~ k_2 = h^2_{I},~ k_3 = h^3_{I},  
\end{align*}

\vskip\baselineskip
\noindent
{\it Case-2}:  $d = 0$. The general solution of Eq.~(\ref{eq:MX=0}), When $d=0$   can be written as 
\begin{equation}
\tilde{X}_1= \tilde{ \alpha}_1\, h^3_{R} +\tilde{ \beta}_1\, h^2_{I} + \tilde{ \gamma}_1\, h^3_{I}
\end{equation}
where $\tilde{\alpha}_1$, $\tilde{\beta}_1$, and $\tilde{\gamma}_1$, are the particular solutions.
The particular solution $\tilde{\alpha}_1$ remains the same as $\alpha_1$. The particular solution $\tilde{\beta}_1$ differs from $\beta_1$, such that $\tilde{\pmb{\beta}}_{R,1}=0$ and $\pmb{\beta}_{I,1}=\tilde{\pmb{\beta}}_{I,1}$. 
 Similarly $\tilde{\pmb{\gamma}}_{R,1}=0$, and $\pmb{\gamma}_{I,1}=\tilde{\pmb{\gamma}}_{I,1}$.  The $2\times2$ matrices corresponding to $\tilde{\beta}_1$ and $\tilde{\gamma}_1$ is PT-Symmetric- normal and belong to $S_2$. The general matrix $\Tilde{H}_1$ in this case can be obtained by fixing $d = 0$ in $H_1$. We can obtain similar structures for the solution spaces of Eq.~(\ref{eq:MX=0}), for all other sets $G_c$. 

Therefore we shall avoid similar discussions in the subsequent cases of  $G_c$, but for the sake of completeness, we shall state the general solutions for each of the cases.

\subsubsection{$G_{s,1}$ -Pseudo Hermitian Matrices}
The set $G_{s,1}$ is characterised by the vector 
\\$\pmb{g}_{R,1}$ and $d = \pm \left|\pmb{g}_{R,1}\right| $ . The general solution of Eq.~(\ref{eq:MX=0}) for this case can be written as 
\begin{equation}\label{eq:X`1}
X'_1 =  \alpha'_1\, h^3_{R} + \beta'_1\, h^1_{I}  + \gamma'_1\,  h^2_{I}   +    \theta'_1\, h^3_{I}
\end{equation}
The particular solutions $\alpha'_1$, $\beta'_1$, $\gamma'_1$ and $\theta'_1$ are given as follows,

\begin{align}
\begin{aligned}
\pmb{\alpha}'_{1} &= \left(\frac{a}{c}, \frac{b}{c}, 1, 0, 0, 0\right);~\quad \pmb{\beta}'_{1} =\left(\frac{a \, b}{ c \,d},\frac{b^2 + c^2}{c \, d}, 0, 1, 0, 0\right);
\\\pmb{\gamma}'_{1}& = \left(-\frac{a^2 + c^2}{c\,d},-\frac{a \,b}{c \, d}, 0, 0, 1, 0\right);~\quad \pmb{\theta}'_{1} = \left(\frac{b}{d},-\frac{a}{d},0,0,0,1\right)& 
\end{aligned}
\end{align}

By using the second condition of  Eq.~(\ref{eq:g0Rh0I})  $h^0_I$ can be obtained from Eq.~(\ref{eq:X`1}). The general $G'_1$-pseudo Hermitian matrix $H'_1$ is given by 
{
\renewcommand{\arraystretch}{1.3}
\begin{equation}\label{eq:H1'}
H'_1=\left[\begin{array}{cc}
\phi_{R,1} - i\phi_{I,1} & \phi_{R,2} - i\phi_{I,2} \\\phi_{R,3}+i\phi_{I,3} & \phi_{R,4} - i\phi_{I,4} \end{array}\right]
\end{equation}}
where,
\begin{align*}
\phi_{R,1} &= m_0 + m_1 ,~
\phi_{I,1} =  m_2\,\frac{a}{d} + m_3\,\frac{b}{d} + m_4\,(\frac{c}{d} - 1),~ 
\phi_{R,2} = m_1\, \frac{ a}{c} +  m_2\,\frac{ a \,b}{c\, d} + m_3 (1-\frac{a^2 + c^2}{c\, d}) + m_4\,\frac{b}{d}, 
\\
\phi_{I,2} &=   m_1\,\frac{b}{c} + m_2\,(\frac{b^2 + c^2}{c\,d}-1) - m_3\,\frac{a\,b}{c\,d} - m_4\,\frac{a}{d},~
\phi_{R,3} =  m_1\,\frac{a}{c} + m_2\,\frac{  a\,b}{c\,d} - m_3\, (1 + \frac{a^2 + c^2}{c\,d}) + m_4\,\frac{b}{d}, 
\\
\phi_{I,3} &=  m_1\,\frac{b}{c} + m_2\,(\frac{b^2 + c^2}{c\,d}+1) - m_3\,\frac{a\,b}{c\,d} - m_4\,\frac{a}{d},~ 
\phi_{R,4} = m_0 -m_1,
\\
\phi_{I,4} &=  m_2\,\frac{a}{d} + m_3\, \frac{b}{d} + m_4\,(\frac{c}{d} + 1),~
 m_0=h^0_R,~ m_1=h^3_R,~ m_2=h^1_I,~ m_3=h^2_I,~ m_4=h^3_I.
\end{align*}
The matrix corresponding to $\alpha'_1$ is PT-symmetric and belongs to $S_1$, while those corresponding to  $\beta'_1$, $\gamma'_1$ and $\theta'_1$ are  not. Hence $H'_1$ is not PT-symmetric in general. When $m_4=-\frac{m_2\,a + m_3\,b}{c}$, then all $H'_1$ given  by Eq.~(\ref{eq:H1'}) depend on 4 parameters, become PT-symmetric and they are all $G_1$-pseudo Hermitian for all $ d \neq\pm\left|\pmb{g}_{R,1}\right|$ (This can be seen in a straight forward manner by putting $m_1 = K_1$; $m_2 = -\frac{(c k_3 + b k_2)}{a}$ and $ m_3= k_2$) .
\subsection{$G_2$-Pseudo Hermitian Matrices}
For this case $\pmb{g}_{R,2}$ = $( a = 0, b \neq0, c\neq0)$ and   $ d \neq \pm|\pmb{g}_{R,2}| $.
\begin{equation}
 G_2=\left\lbrace \left[\begin{array}{cc}
d + c & -i b \\
i b & d - c
\end{array}\right]: b,c\in\mathbb{R}^{*}, d\in\mathbb{R}\right\rbrace
\end{equation}

\noindent
{\it Case-1: $d\neq0$}. The general solution of Eq.~(\ref{eq:MX=0}) for this case can be written as 
\begin{equation}
X_2 =  \alpha_2\, h^3_{R} + \beta_2\, h^1_{I} + \gamma_2\, h^3_{I}
\end{equation}
where $ h^3_{R}, h^1_{I}, h^3_{I}\in\mathbb{R} $ . The particular solutions $\alpha_2$, $\beta_2$, $\gamma_2$ are given as follows   .

\begin{eqnarray}
\pmb{\alpha}_{2} = \left(0,\frac{b}{c}, 1, 0, 0, 0\right);\quad\pmb{\beta}_{2} = \left(0,\frac{d}{c}, 0, 1, 0, 0\right);\quad\pmb{\gamma}_{2} = \left(\frac{d}{b}, 0, 0, 0, -\frac{c}{b}, 1\right) 
\end{eqnarray}

The general $G_2$-pseudo Hermitian matrix $H_2$ is given by 
{
\renewcommand{\arraystretch}{1.3}
\begin{equation}
H_2=\left[\begin{array}{cc}
k_0 + k_1 + i k_3   & k_3\,\frac{(d -c)}{b}  - i( k_1\, \frac{b}{c} - k_2\,(1- \frac{d}{c}))\\ k_3\,\frac{ (c + d)}{b}  + i( k_1\,\frac{b}{c} + k_2\,(1+ \frac{d}{c}))  &  k_0  - (k_1 + i k_3) \end{array}\right]
\end{equation}}
where  $k_0 = h^0_{R}$, $k_1 = h^3_{R}$, $k_2 = h^1_{I}$, $k_3 = h^3_{I}$,  
\vskip\baselineskip
\noindent
{\it Case-2:  $d = 0$}. The general solution for this case can be written as 
\begin{equation}
\tilde{X}_2 =  \tilde{\alpha}_2\, h^3_{R} + \tilde{\beta}_2\, h^2_{I} + \tilde{\gamma}_1\, h^3_{I}
\end{equation}
The general matrix $\Tilde{H}_2$ in this case can be obtained by fixing $d = 0$ in $H_2$.
\subsubsection{$G_{s,2}$-Pseudo Hermitian Matrices}
The set $G_{s,2}$ is characterised by the vector 
\\$\pmb{g}_{R,2}$ and $d = \pm\left|\pmb{g}_{R,2}\right| $ . The general solution for this case can be written as 
\begin{equation}
X'_2 = \alpha'_2 \, h^3_{R}   + \beta'_2 \, h^1_{I}   + \gamma'_2 \, h^2_{I}   +  \theta'_2 \, h^3_{I}  
\end{equation}
The particular solutions $\alpha'_2$ ,$\beta'_2$, $\gamma'_2$ and $\theta'_2$ are given as follows,

\begin{align}
\begin{aligned}
\pmb{\alpha}'_{2} = \left(0,\frac{b}{c},1,0,0,0\right);~\quad \pmb{\beta}'_{2} = \left(0,\frac{d}{c}, 0, 1, 0, 0\right);
\\
\pmb{\gamma}'_{2} = \left(-\frac{c}{d},0, 0,0, 1, 0\right);~\quad
\pmb{\theta}'_{2} = \left(\frac{b}{d},0,0,0,0,1\right) 
\end{aligned}
\end{align}

The general $G_{s,2}$-pseudo Hermitian matrix $H'_2$ is given by 
{
\renewcommand{\arraystretch}{1.5}
\begin{equation}\label{eq:H2'}
H'_2=\left[\begin{array}{cc}
m_0 + m_1 - i(m_3\,\frac{b}{d} + m_4 \,(\frac{c}{d} - 1) ) &  m_3\, (1- \frac{c}{d}) + m_4\,\frac{b}{d}  - i\phi_1\\ -m_3 \,(1 + \frac{c}{d}) + m_4\,\frac{b}{d}  + i\phi_2  & m_0 - m_1 - i(m_3\,\frac{b}{d} + m_4 \,(\frac{c}{d} + 1) )\end{array}\right]
\end{equation}}
where 
\begin{align*}
    \phi_1 = m_1\,\frac{b}{c} -  m_2\, (1- \frac{d}{c}),~  
 \phi_2 = m_1\,\frac{b}{c} +  m_2\, (1 + \frac{d}{c}) ,\\
m_0 = h^0_{R},~ m_1 = h^3_{R},~ m_2 = h^1_{I},~ m_3 = h^2_{I},~ m_4 = h^3_{I}
\end{align*}

Note that matrices corresponding to   $\alpha'_2$ and $\beta'_2$ are  PT-symmetric and belong to $S_1$ and $S_4$ respectively, while the matrices  corresponding to $\gamma'_2$ and $\theta'_2$ are not. Hence $H'_2$ is not PT-symmetric in general. When $ m_4 = -m_3\,\frac{b}{c}$ then all $H'_2$ given  by Eq.~(\ref{eq:H2'})  become PT-symmetric and they are all $G_2$-pseudo Hermitian for  all $d \neq \pm\left|\pmb{g}_{R,2}\right| $. 

\subsection{$G_3$-Pseudo Hermitian Matrices}For this case $\mathbf{g}_{R,3}$ = $\left(a\neq0, b=0, c\neq0\right)$ and   $ d \neq \pm|\pmb{g}_{R,3}| $.
\begin{equation}
 G_3=\left\lbrace\left[\begin{array}{cc}
d + c  & a \\
a & d - c
\end{array}\right]:a,c\in\mathbb{R}^{*}, d \in \mathbb{R}\right\rbrace
\end{equation}

\noindent
{\it Case-1: $d\neq0$}. The general solution of Eq.~(\ref{eq:MX=0}) for this case can be written as 
\begin{equation}
X_3 =  \alpha_3\, h^3_{R} + \beta_3\, h^2_{I} + \gamma_3\, h^3_{I}
\end{equation}
where $h^3_{R,3},h^2_{I,3},h^3_{I,3}\in\mathbb{R}$. The particular solutions are given as follows,

\begin{eqnarray}
\pmb{\alpha}_{3} = \left(\frac{a}{c},0,1,0,0,0\right);\quad
\pmb{\beta}_{3} = \left(-\frac{d}{c},0, 0,0,1,0\right);\quad\pmb{\gamma}_{3} = \left(0,-\frac{d}{a},0,-\frac{c}{a},0, 1\right)
\end{eqnarray}

The general $G_3$-pseudo Hermitian matrix $H_3$ is given by
{
\renewcommand{\arraystretch}{1.5}
\begin{equation}
H_3=\left[\begin{array}{cc}
k_0 + k_1 + i k_3   & k_1\,\frac{a}{c} + k_2\,(1 -\frac{d}{c}) + ik_3\,\frac{(d -c)}{a}) \\ k_1\,\frac{a}{c} -k_2\,(1 +\frac{d}{c}) - ik_3\,\frac{(d+c)}{a})  &  k_0 - (k_1 + i k_3) \end{array}\right]
\end{equation}}
where  $k_0=h^0_{R}$, $k_1=h^3_{R}$, $k_2 = h^2_{I}$, $k_3 = h^3_{I}$, 
\vskip\baselineskip
\noindent
{\it Case-2: $d = 0$}. The general solution for this case can be written as 
\begin{equation}
\tilde{X}_3 =  \tilde{\alpha}_3\, h^3_{R} + \tilde{\beta}_3\, h^2_{I} + \tilde{\gamma}_3\, h^3_{I}
\end{equation}
The general matrix $\Tilde{H}_3$ in this case can be obtained by fixing $d = 0$ in $H_3$.
\subsubsection{$G_{s,3}$-Pseudo Hermitian Matrices}The set $G_{s,3}$ is characterised by 
 the vector 
\\$\pmb{g}_{R,3}$ = $(a\neq0, b = 0, c\neq0)$
and $d = \pm\left|\pmb{g}_{R,3}\right| $ . The general solution for this case can be written as 
\begin{equation}
X'_3 = \alpha'_3 \, h^3_{R}  + \beta'_3\, h^2_{I}  + \gamma'_3\,  h^1_{I}   + \theta'_3\, h^3_{I}  
\end{equation}
The particular solutions $\alpha'_3$ ,$\beta'_3$, $\gamma'_3$ and $\theta'_3$ are given as follows,
\begin{align}
\begin{aligned}
\pmb{\alpha}'_{3} &= \left(\frac{a}{c},0,1,0,0,0\right);~\quad
\pmb{\beta}'_{3} = \left(-\frac{d}{c},0, 0,0, 1, 0\right);
\\
\pmb{\gamma}'_{3} &= \left(0,\frac{c}{d}, 0, 1, 0, 0\right);~\quad\pmb{\theta}'_{3} = \left(0,-\frac{a}{d},0,0,0,1\right) 
\end{aligned}
\end{align}

The general $G_{s,3}$-pseudo Hermitian matrix $H'_3$ is given by 
{
\renewcommand{\arraystretch}{1.5}
\begin{equation}\label{eq:H3'}
H'_3=\left[\begin{array}{cc}
m_0 + m_1 - i(m_3\,\frac{a}{d} + m_4\,(\frac{c}{d} -  1) ) &  m_1\,\frac{a}{c} + m_2\, (1 - \frac{d}{c})   + i\phi_1\\   m_1\,\frac{a}{c} - m_2\, (1 + \frac{d}{c})  + i\phi_2  & m_0 - m_1 - i(m_3\,\frac{a}{d} + m_4\,(\frac{c}{d} +  1) )\end{array}\right]
\end{equation}}
where\begin{align*}
\phi_1 = m_3\, (1- \frac{c}{d}) + m_4\,\frac{ a}{d} ,~ 
 \phi_2 = m_3\, (1+ \frac{c}{d}) - m_4\,\frac{ a}{d} ,~
 m_0 = h^0_{R},~ m_1  =h^3_{R},~
m_2 = h^2_{I},~ m_3 = h^1_{I},~ m_4 = h^3_{I}.\end{align*}
Note that the matrices corresponding to $\alpha'_3$ and $\beta'_3$  are  PT-symmetric and belong to $S_1$ and $S_4$ respectively, while those corresponding to $\gamma'_3$ and $\theta'_3$ are not. Hence $H'_3$ is not PT-symmetric in general. When $ m_4 = -m_3\,\frac{a}{c}$ then all $H'_3$ given  by Eq.~(\ref{eq:H3'}) depend on 4 parameters, become PT-symmetric and they are all  $G_3$-pseudo Hermitian for all $d \neq\pm \left|\pmb{g}_{R,3}\right| $  . 

\subsection{$G_4$-Pseudo Hermitian Matrices} For this case $\pmb{g}_{R,4}$ = $(a\neq0,  b\neq0,  c = 0)$ and  $ d \neq\pm |\pmb{g}_{R,4}| $.
\begin{equation}
 G_4=\left\lbrace\left[\begin{array}{cc}
d  & a - ib \\
a + ib & d
\end{array}\right]: a,b \in \mathbb{R}^{*}, d\in\mathbb{R}\right\rbrace
\end{equation}
\\{\it Case-1:  $d\neq0$}. The general solution of Eq.~(\ref{eq:MX=0}) for this case can be written as 
\begin{equation}
X_4 =  \alpha_4\, h^2_{R} + \beta_4\, h^2_{I} + \gamma_4\, h^3_{I}
\end{equation}
where $ h^2_{R}, h^2_{I}, h^3_{I}\in\mathbb{R}$ . The particular solutions $\alpha_4$ , $\beta_4$, and $\gamma_4$ are given as follows,

\begin{eqnarray}
\pmb{\alpha}_{4} = \left(\frac{a}{b},1,0,0,0,0\right);\quad \pmb{\beta}_{4} = \left(0,0,\frac{d}{a},-\frac{b}{a}, 1, 0\right);\quad \pmb{\gamma}_{4} = \left( \frac{d}{b},0, 0,0, 0, 1\right)
\end{eqnarray}
The general $G_4$-pseudo Hermitian matrix $H_4$ is given by
{
\renewcommand{\arraystretch}{1.5}
\begin{equation}
H_4=\left[\begin{array}{cc}
k_0 + k_2\,\frac{d}{a} + i k_3   & k_1\,\frac{a}{b} + k_2 + k_3\,\frac{d}{b} - i(k_1 + k_2\,\frac{b}{a}) \\ k_1\,\frac{a}{b} - k_2 + k_3\,\frac{d}{b} + i(k_1 - k_2\,\frac{b}{a}) &  k_0 - ( k_2\,\frac{d}{a} + i k_3)  \end{array}\right]
\end{equation}}
where  $k_0 = h^0_{R}$, $k_1 = h^2_{R}$, $k_2 = h^2_{I}$, $k_3 = h^3_{I}$, 

\vskip\baselineskip
\noindent
{\it Case-2:  $d = 0$}. The general solution for this case can be written as 
\begin{equation}
\tilde{X}_4 =  \tilde{\alpha}_4\, h^2_{R} + \tilde{\beta}_4\, h^2_{I} + \tilde{\gamma}_4\, h^3_{I}
\end{equation}
The general matrix $\Tilde{H}_4$ in this case can be obtained by fixing $d = 0$ in $H_4$.
\subsubsection{$G_{s,4}$-Pseudo Hermitian Matrices}The set $G_{s,4}$ is characterised by 
 the vector 
\\$\pmb{g}_{R,4}$ and $d = \pm \left|\pmb{g}_{R,4}\right| $ . The general solution for this case can be written as 
\begin{equation}
X'_4 = \alpha'_4\,  h^2_{R}  + \beta'_4\, h^1_{I}   + \gamma'_4 \, h^3_{I}   +  \theta'_4 \,h^2_{I}  
\end{equation}
The particular solutions $\alpha'_4$ ,$\beta'_4$, $\gamma'_4$ and $\theta'_4$ are given as follows,

\begin{align}
\begin{aligned}
\pmb{\alpha}'_{4} &= \left(\frac{a}{b},1,0,0,0,0\right);~\quad \pmb{\beta}'_{4} = \left(0,0,-\frac{b}{d},1, 0, 0\right);
\\
 \pmb{\gamma}'_{4} &= \left(\frac{d}{b},0,0,0,0,1\right);~\quad\pmb{\theta}'_{R,4} = \left(0, 0,\frac{a}{d},0, 1, 0\right)  
\end{aligned}
\end{align}

The general $G'_4$-pseudo Hermitian matrix $H'_4$ is given by 
{
\renewcommand{\arraystretch}{1.5}
\begin{equation}\label{eq:H4'}
H'_4=\left[\begin{array}{cc}
m_0 - m_2\,\frac{b}{d} + m_4\,\frac{a}{d}  - i\phi_1 &  m_1\,\frac{a}{b} + m_3\,\frac{d}{b} + m_4   + i\phi_2\\   m_1\,\frac{a}{b} + m_3\,\frac{d}{b} - m_4   + i\phi_3  & m_0 + m_2\,\frac{b}{d} - m_4\,\frac{a}{d}  - i\phi_4\end{array}\right]
\end{equation}}
where
\begin{align*}
    \phi_1 &= m_2\,\frac{a}{d} - m_3 + m_4\,\frac{b}{d}  ,~ \phi_2 = m_2 -m_1 ,~ \phi_3 = m_1 + m_2 ,~  
 \phi_4 = m_2\,\frac{a}{d} + m_3 + m_4\,\frac{b}{d} ,
 \\
m_0 &= h^0_{R},~ m_1 = h^2_{R},~ m_2 = h^1_{I},~ m_3 = h^3_{I}, ~ m_4 = h^2_{I}.
\end{align*}
Note that the matrices corresponding to  $\alpha'_4$ and $\gamma'_4$   are  PT-symmetric and belong to $S_1$ and $S_4$ while those corresponding to $\beta'_4$ and $\theta'_4$  are not. Hence $H'_4$ is not PT-symmetric in general. When $m_4 = -m_2\,\frac{a}{b}$ then, all $H'_4$ given by Eq.~(\ref{eq:H4'}) depend on 4 parameters, become PT-symmetric and they become $G_4$-pseudo Hermitian for all $d \neq \pm\left|\pmb{g}_{R,4}\right| $. 

\subsection{$G_5$-Pseudo Hermitian Matrices}
For this case $\pmb{g}_{R,5}$ = $\left(  a \neq0,  b = 0,  c = 0 \right)$ and  $ d \neq \pm|\pmb{g}_{R,5}| $.
\begin{equation}
 G_5 = \left\lbrace\left[\begin{array}{cc}
d  & a \\
a  & d 
\end{array}\right]:a \in\mathbb{R}^{*},d\in \mathbb{R} \right\rbrace
\end{equation}
\\{\it Case-1: $d\neq0$}. The general solution of Eq.~(\ref{eq:MX=0}) for this case can be written as 
\begin{equation}
X_5 =  \alpha_5\, h^1_{R} + \beta_5\, h^2_{I} + \gamma_5\, h^3_{I}
\end{equation}
where $ h^1_{R}, h^2_{I}, h^3_{I}\in\mathbb{R}$ . The particular solutions $\alpha_5$ , $\beta_5$, and $\gamma_5$ are given as follows,
\begin{eqnarray}
\pmb{\alpha}_{5} = \left(1,0,0,0,0,0\right);\quad 
\pmb{\beta}_{5} = \left(0,0,\frac{d}{a},0,1, 0\right);\quad
\pmb{\gamma}_{5} = \left(0,-\frac{d}{a},0,0,0, 1\right)
\end{eqnarray}
The general $G_5$-pseudo Hermitian matrix $H_5$ is given by
{
\renewcommand{\arraystretch}{1.5}
\begin{equation}
H_5=\left[\begin{array}{cc}
k_0 + k_2\,\frac{d}{a} + i k_3   & k_1 + k_2 + ik_3\,\frac{d}{a}\\ k_1 - k_2 - ik_3\,\frac{d}{a} &  k_0 -( k_2\,\frac{d}{a} + i k_3)  \end{array}\right]
\end{equation}}
where  $k_0 = h^0_{R}$, $k_1 = h^1_{R}$, $k_2 = h^2_{I}$, $k_3 = h^3_{I}$, 

\vskip\baselineskip
\noindent
{\it Case-2:  $d= 0$}. The general solution for this case can be written as 
\begin{equation}
\tilde{X}_5 =  \tilde{\alpha}_5\, h^1_{R} + \tilde{\beta}_5\, h^2_{I} + \tilde{\gamma}_5\, h^3_{I}
\end{equation}
The general matrix $\Tilde{H}_5$ in this case can be obtained by fixing $d = 0$ in $H_5$.
\subsubsection{$G_{s,5}$-Pseudo Hermitian Matrices}The set $G_{s,5}$ is characterised by 
 the vector 
\\$\pmb{g}_{R,5}$ and $| =\pm \left|\pmb{g}_{R,5}\right| $. The general solution for this case can be written as 
\begin{equation}
X'_5 = \alpha'_5\, h^1_{R}   +  \beta'_5 \, h^2_{I} + \gamma'_5\, h^3_{I}   +  \theta'_5\, h^1_{I}  
\end{equation}
The particular solutions $\alpha'_5$, $\beta'_5$, $\gamma'_5$ and $\theta'_5$ are given as follows,

\begin{align}
\begin{aligned}
\pmb{\alpha}'_{5} &= \left(1,0,0,0,0,0\right);~\quad 
\pmb{\beta}'_{5} = \left(0, 0,\frac{d}{a},0, 1, 0\right); 
\\
\pmb{\gamma}'_{5} &= \left(0,-\frac{d}{a},0,0,0,1\right);~\quad\pmb{\theta}'_{5} = \left(0,0,0,1, 0, 0\right)
\end{aligned}
\end{align}

The general $G_{s,5}$-pseudo Hermitian matrix $H'_5$ is given by 
{
\renewcommand{\arraystretch}{1.5}
\begin{equation}\label{eq:H5'}
H'_5=\left[\begin{array}{cc}
m_0 + m_2\,\frac{ d}{a}  - i(m_4\,\frac{d}{a} -  m_3 ) & m_1 + m_2  + i(m_4 + m_3\,\frac{d}{a})\\ m_1 - m_2  + i(m_4 - m_3\,\frac{d}{a})   &  m_0 - m_2\,\frac{d}{a}  - i(m_4\,\frac{d}{a} +  m_3 ))\end{array}\right]
\end{equation}}
where, $m_0 = h^0_{R}$,~ $m_1 = h^1_{R}$,~ $m_2 = h^2_{I}$,~ $m_3 = h^3_{I}$,~ $m_4 = h^1_{I}$.

Note that the matrix  corresponding to  $\alpha'_5$ belongs to $S_1$, and matrices corresponding to $\beta'_5$, and $\gamma'_5$  are   belonging to  $S_4$. The matrix corresponding to $\theta'_5$  is not PT-symmetric. Hence $H'_5$ is not PT-symmetric in general. When $m_4=0$, then all $H'_5$ given by Eq.~(\ref{eq:H5'}) depend on 4 parameters, become PT-symmetric and they are all  $G_5$-Pseudo Hermitian for all $d \neq \pm\left|\pmb{g}_{R,5}\right| $. 
\subsection{$G_6$-Pseudo Hermitian Matrices}
For this case $\pmb{g}_{R,6}$ = $\left( a = 0,  b \neq0,  c = 0 \right)$ and $ d \neq \pm|\pmb{g}_{R,6}| $.
\begin{equation}
 G_6=\left\lbrace\left[\begin{array}{cc}
d  & -ib \\
ib & d
\end{array}\right]:b\in\mathbb{R}^{*},d\in \mathbb{R}\right\rbrace
\end{equation}
\\{\it Case-1:  $d\neq0$}. The general solution of Eq.~(\ref{eq:MX=0}) for this case can be written as 
\begin{equation}
X_6 =  \alpha_6\, h^2_{R} + \beta_6\, h^1_{I} + \gamma_6\, h^3_{I}
\end{equation}
where $ h^2_{R}, h^1_{I}, h^3_{I}\in\mathbb{R}$ . The particular solutions $\alpha_6$, $\beta_6$, and $\gamma_6$ are given as follows,
\begin{eqnarray}
\pmb{\alpha}_{6} = \left(0,1,0,0,0,0\right);\quad \pmb{\beta}_{6} = \left(0,0,-\frac{d}{b},1,0, 0\right);\quad
\pmb{\gamma}_{6} = \left(\frac{d}{b},0,0,0,0, 1\right)
\end{eqnarray}
The general $G_6$-pseudo Hermitian matrix $H_6$ is given by
{
\renewcommand{\arraystretch}{1.5}
\begin{equation}
H_6 =\left[\begin{array}{cc}
k_0 -k_2\,\frac{d}{b} + i k_3   &  k_3\,\frac{d}{b} -i (k_1 - k_2)  \\ k_3\,\frac{d}{b} + i (k_1 + k_2) &  k_0 + k_2\, \frac{d}{b} -i k_3  \end{array}\right]
\end{equation}}
where  $k_0 = h^0_{R}$, $k_1 = h^2_{R}$, $k_2 = h^1_{I}$, $k_3 = h^3_{I}$,
\vskip\baselineskip
\noindent
{\it Case-2: $d = 0$}. The general solution for this case can be written as 
\begin{equation}
\tilde{X}_6 =  \tilde{\alpha}_6\, h^2_{R} + \tilde{\beta}_6\, h^1_{I} + \tilde{\gamma}_6\, h^3_{I}
\end{equation}
The general matrix $\Tilde{H}_6$ in this case can be obtained by fixing $d = 0$ in $H_6$.
\subsubsection{$G_{s,6}$-Pseudo Hermitian Matrices}
The set $G_{s,6}$ is characterised by 
 the vector 
\\$\pmb{g}_{R,6}$ and $d = \pm\left|\pmb{g}_{R,6}\right| $ . The general solution for this case can be written as 
\begin{equation}
X'_6 = \alpha'_6\,  h^2_{R}  + \beta'_6\, h^1_{I}  + \gamma'_6 \, h^3_{I}   +  \theta'_6\, h^2_{I}  
\end{equation}
The particular solutions $\alpha'_6$ ,$\beta'_6$, $\gamma'_6$ and $\theta'_6$ are given as follows,

\begin{align}
\begin{aligned}
\pmb{\alpha}'_{6} &= \left(0,1,0,0,0,0\right);
~\quad\quad\pmb{\beta}'_{6} = \left(0,0,-\frac{d}{b},1, 0, 0\right);
\\
\pmb{\gamma}'_{6} &= \left(\frac{d}{b},0,0,0,0,1\right); 
~\quad\pmb{\theta}'_{R,6} =  \left(0, 0,0,0, 1, 0\right)
\end{aligned}
\end{align}

The general $G_{s,6}$-pseudo Hermitian matrix $H'_6$ is given by 
{
\renewcommand{\arraystretch}{1.5}
\begin{equation}\label{eq:H6'}
H'_6=\left[\begin{array}{cc}
m_0 - m_2\, \frac{d}{b}  - i(-m_3 + m_4\,\frac{ d }{b} ) & m_3\, \frac{ d}{b} + m_4   - i(m_1 - m_2)\\ m_3\, \frac{ d}{b} - m_4   +i(m_1 + m_2) & m_0 + m_2\, \frac{d}{b}  - i(m_3 + m_4\, \frac{ d }{b}   )\end{array}\right]
\end{equation}}
where, $m_0 = h^0_{R}$,~ $m_1 = h^2_{R}$,~ $m_2 = h^1_{I}$,~ $m_3 = h^3_{I}$,~ $m_4 = h^2_{I}$.

Note that the matrix corresponding to  $\alpha'_6$ belongs to $S_1$, and the matrices corresponding to  $\beta'_6$ and $\gamma'_6$ are  belonging to $S_4$. The matrix corresponding to  $\theta'_6$ is not PT-symmetric. Hence $H'_6$ is not PT-symmetric in general. When $m_4=0$ then all $H'_6$ given by Eq.~(\ref{eq:H6'}) depend on 4 parameters, become PT-symmetric and they are all  $G_6$-pseudo Hermitian for all $d \neq \pm\left|\pmb{g}_{R,6}\right| $. 
\subsection{$G_7$-Pseudo Hermitian Matrices}
For this case $\pmb{g}_{R,7}$ = $\left( a = 0,  b = 0,  c\neq0 \right)$ and  $ d \neq \pm|\pmb{g}_{R,7}| $.
\begin{equation}
 G_7=\left\lbrace\left[\begin{array}{cc}
d + c  & 0 \\
0 & d - c
\end{array}\right]:c\in\mathbb{R}^{*},d\in \mathbb{R}\right\rbrace
\end{equation}
\\{\it Case-1: $d\neq0$}. The general solution of Eq.~(\ref{eq:MX=0}) for this case can be written as 
\begin{equation}
X_7 =  \alpha_7\, h^3_{R} + \beta_7\, h^1_{I} + \gamma_7\, h^2_{I}
\end{equation}
where $ h^3_{R}, h^1_{I}, h^2_{I}\in\mathbb{R}$ . The particular solutions $\alpha_7$, $\beta_7$, and $\gamma_7$ are given as follows,
\begin{eqnarray}
\pmb{\alpha}_{7} =  \left(0,0,1,0,0,0\right);\quad
 \pmb{\beta}_{7} = \left(0,\frac{d}{c},0,1,0, 0\right);\quad
\pmb{\gamma}_{7} = \left(-\frac{d}{c},0,0,0,1, 0\right)
\end{eqnarray}
The general $G_7$-pseudo Hermitian matrix $H_7$ is given by
{
\renewcommand{\arraystretch}{1.5}
\begin{equation}
H_7=\left[\begin{array}{cc}
k_0 + k_1    & k_3\,(1 -\frac{d}{c})+i k_2\,(1 -\frac{d}{c}) \\  -k_3\,(1 +\frac{d}{c})+i k_2\,(1 +\frac{d}{c}) &  k_0  - k_1 \end{array}\right]
\end{equation}}
where  $k_0 = h^0_{R}$, $k_1 = h^3_{R}$, $k_2 = h^1_{I}$, $k_3 = h^2_{I}$,
\vskip\baselineskip
\noindent
{\it Case-2:  $d = 0$}. The general solution for this case can be written as 
\begin{equation}
\tilde{X}_7 =  \tilde{\alpha}_7\, h^3_{R} + \tilde{\beta}_7\, h^1_{I} + \tilde{\gamma}_7\, h^2_{I}
\end{equation}
The general matrix $\Tilde{H}_7$ in this case can be obtained by fixing $d = 0$ in $H_7$.
\subsubsection{$G_{s,7}$-Pseudo Hermitian Matrices}The set $G_{s,7}$ is characterised by 
 the vector 
\\$\pmb{g}_{R,7}$ and $d = \pm\left|\pmb{g}_{R,7}\right|$ . The general solution for this case can be written as 
\begin{equation}
X'_7 = \alpha'_7\, h^3_{R}  + \beta'_7\,   h^1_{I} +  \gamma'_7 \, h^2_{I}   +\theta'_7\, h^3_{I}  
\end{equation}

The particular solutions $\alpha'_7$, $\beta'_7$, $\gamma'_7$ and $\theta'_7$ are given as follows,

\begin{align}
\begin{aligned}
\pmb{\alpha}'_{7}& = \left(0,0,1,0,0,0\right);~\quad
\pmb{\beta}'_{7} =  \left(0,\frac{d}{c},0,1, 0, 0\right);\\
\pmb{\gamma}'_{7}& =  \left(-\frac{d}{c}, 0,0,0, 1, 0\right);~\quad
\pmb{\theta}'_{7}= \left(0,0,0,0,0,1\right)
\end{aligned}
\end{align}

The general $G'_7$-pseudo Hermitian matrix $H'_7$ is given by 
{
\renewcommand{\arraystretch}{1.5}
\begin{equation}\label{eq:H7'}
H'_7=\left[\begin{array}{cc}
m_0 + m_1  + im_4\, (1 -\frac{d}{c} ) & m_3\, (1 - \frac{d}{c}) + im_2\, (1 - \frac{d}{c})\\- m_3\, (1 + \frac{d}{c}) + im_2\, (1 + \frac{ d}{c})& m_0 - m_1  - im_4\, (1 +\frac{d }{c} ))\end{array}\right]
\end{equation}}
where, ~$m_0=h^0_{R,7}$, ~$m_1=h^3_{R,7}$, ~$m_2=h^1_{I,7}$, ~$m_3=h^2_{I,7}$, ~$m_4=h^3_{I,7}$.

Note that the matrix corresponding to  $\alpha'_7$ belongs to $S_1$,  and the matrices corresponding to $\beta'_7$ and $\gamma'_7$ are belonging to $S_4$.  The matrix corresponding to  $\theta'_7$ is not PT-symmetric. Hence $H'_7$ is not PT-symmetric in general. When $ m_4=0$
 then all $H'_7$ given by Eq.~(\ref{eq:H7'}) depend on 4 parameters, become PT-symmetric and they are all  $G_7$-pseudo Hermitian for all $d \neq\pm\left|\pmb{g}_{R,7}\right| $. 

 The above discussion on the solution spaces corresponding to Eq.~(\ref{eq:MX=0}) for each $G_c(G_s,c)$ leads to the following general observations:
\begin{enumerate}
\item For a given invertible $G$ such that $Trace(G) = 0$, the set of all admissible $H$ satisfying Eq.~(\ref{eq:HdaggerG}) forms a 
four-dimensional vector space. All such trace-less $H$ can be written as a linear combination of three normal
matrices, associated with $\tilde{\alpha},\tilde{\beta},\tilde{\gamma}$ where the matrix associated with $\tilde{\alpha}$  belongs to $S_1$ and matrices associated with $\tilde{\beta},\tilde{\gamma}$ belong to $S_2$.\\
\item For a given  invertible $G$ such that $Trace(G) \neq0$, the set  of all admissible $H$
satisfying Eq.~(\ref{eq:HdaggerG}) forms a four-dimensional vector space. All such trace-less $H$ can be written as a linear combination
of one normal matrix belongs to $S_1$ and two non-normal matrices belong to $\in S_4$, associated with $\alpha,\beta,\gamma$ respectively.\\
\item For a given singular $G_s$ the set of all $G_s$-pseudo Hermitian matrices with $h^0_R = 0$, forms a vector space of dimension four, such that the set of all PT-symmetric $G_s$-pseudo Hermitian matrices forms a three dimensional subspace.
\end{enumerate}
These observations lead to the following theorem:
\begin{thm}
Let $H\in M_{2}(\mathbb{C})$ and $G\in S_{1}$, If $H^{\dagger}G =GH$, then
\begin{enumerate}
\item for $G$ such that $Trace(G) = 0$, the set of all trace-less $G$-pseudo Hermitian $H$ can be written as the linear combination of one normal PT-symmetric matrix belongs to $S_1$ and  two normal PT-symmetric matrices belong to $S_2$.
\item for  $G$ such that $Trace(G)\neq0$, the set of all trace-less $G$-pseudo Hermitian $H$ can be written 
as linear combination of one normal PT- symmetric matrix  belongs to $S_1$ and two 
non-normal PT-symmetric matrices belong to $S_4$.
\item for singular $G_s$, the set of all trace-less $G_s$-pseudo Hermitian $H$ can be written 
as linear combination of one normal PT- symmetric matrix  belongs to $S_1$ and two 
non-normal PT-symmetric matrices belong to $S_4$.
\end{enumerate}
\end{thm}

\section{Determinant of Ensemble of $G$-Pseudo Hermitian Matrices and Quadrics}\label{sec:5}
As discussed in section~\ref{sec:4}, we get seven ensembles of general $ 2\times 2$  pseudo-Hermitian  matrices corresponding to seven classes of $G_i$ matrices. They are given by $H_i$ (refer to section~\ref{sec:4} for details).
Now we study the determinant of trace-less $H_i$ ($i = 1 \cdot\cdot\cdot 7$), which turns out to
be a quadratic form given by
\begin{equation}
\det(H_i) = v^T A_i v  : A_i = A_i^T \in M_3(\mathbb{R}), \, v^T =(k_1,k_2,k_3)
 \end{equation}
Where $k_1,k_2,k_3\in\mathbb{R}$ are the free variables of the associated ensembles of matrices.
     
We are interested in computing the $\det (H_i)$ for the cases 1) determinant of $G$ is positive,   2) determinant of $G$ is negative.
For $i=5,6,7$ the $A_i$ matrices  are given by
{
\renewcommand{\arraystretch}{1.3}
\begin{equation}
A_5 = \left[\begin{array}{ccc}
-1 &  0 & 0\\
0 & \frac{a^{2} - d^{2}}{a^{2}} & 0\\
0 & 0 & \frac{a^2 - d^2}{a^{2}}
\end{array}\right]
\end{equation}
\begin{equation}
A_6 = \left[\begin{array}{ccc}
-1 &  0 & 0\\
0 & \frac{b^2 - d^2}{b^{2}} & 0\\
0 & 0 & \frac{b^2 - d^2}{b^2}
\end{array}\right]
\end{equation}
\begin{equation}
A_7 = \left[\begin{array}{ccc}
-1 &  0 & 0\\
0 & \frac{c^2 - d^2}{c^2} & 0\\
0 & 0 & \frac{c^2 - d^2}{c^2}
\end{array}\right]
\end{equation}}

 Note that $A_{i}$ for $i=5,6,7$ are diagonal matrices.  
 
 If the determinant of $G_i$ is positive then all the eigenvalues of $A_{i}$ for $i=5,6,7$  are negative and the determinant of the associated ensemble of matrices ($H_i$) become negative and it represents  PT-symmetric matrices with real eigenvalues. Thus when determinant of $G_i$ is positive, then $\det(H_i)=-\lambda :\lambda >0$ (for $i=5,6,7$) 
  defines an ellipsoid. 
 Similarily, if the determinant of $G_i$ is negative then the eigenvalues of $A_{i}$ for $i=5,6,7$ other than -1 become positive and  the determinant of the associated ensemble of matrices ($H_i$) can be positive, negative  and zero. In this case the ensemble of matrices ($H_i$) can represent   PT-symmetric matrices with real eigenvalues  as well as complex eigenvalues . Thus when determinant of $G_i$ is negative,  $\det(H_i)= \lambda$, $\det(H_i)=  0$, $\det(H_i)=  -\lambda :\lambda>0$ (for $i=5,6,7$) 
defines hyperboloid of one sheet, quadric cone, and hyperboloid of two sheets respectively.

 For $i=2,3,4$ the $A_i$ matrices  are given as
{
\renewcommand{\arraystretch}{1.5}
\begin{equation}
A_2 = \left[\begin{array}{ccc}
-\frac{b^2 + c^2}{c^2} & -\frac{b\,d}{c^2} & 0\\
-\frac{b\,d}{c^2} & \frac{c^2 - d^2}{c^2} & 0\\ 
0 & 0 & \frac{b^2 + c^2 - d^2}{b^2}
\end{array}\right]
\end{equation}
\begin{equation}
A_3 = \left[\begin{array}{ccc}
-\frac{a^2 + c^2}{c^2} & \frac{a\,d}{c^2} & 0\\
\frac{a\,d}{c^2} & \frac{c^2 - d^2}{c^2} & 0\\
0 & 0 & \frac{a^2 + c^2 - d^2}{a^2}
\end{array}\right]
\end{equation} 
\begin{equation}
A_4 = \left[\begin{array}{ccc}
-\frac{a^2 + b^2}{b^2} &  0 & -\frac{a\,d}{b^2}\\
0 & \frac{a^2 + b^2 - d^2}{a^2} & 0\\
-\frac{a\,d}{b^2} & 0 & \frac{b^2 - d^2}{b^2}
\end{array}\right]
\end{equation}}

When $d =0$, ($\det(G_i)$ is negative) $A_i$ becomes diagonal matrix (for $i=2,3,4$) and each $A_i$ has one negative and two positive eigenvalues. As discussed earlier  $\det(H_i)= \lambda$, $\det(H_i)= 0$, $\det(H_i)=  -\lambda : \lambda>0 $ (for $i=2,3,4$) also defines hyperboloid of one sheet, quadric cone, and hyperboloid of two sheets respectively.

When $d  \neq0$, $A_i$ are not diagonal matrices (for $i=2,3,4$) then $\det(H_i)$ in standard  quadratic form is given by
\begin{equation}
\det (H_i) = v^T A_i v = \tilde{v}^T D_i \tilde{v}, 
\end{equation}
Where $D_i = P H P^T$ and $\tilde{v} = P^T v$. 
For $i=2,3,4$, $D_i$ matrices are given as,
\begin{equation}
D_2 = \left[\begin{array}{ccc}
\frac{b^2 + c^2 - d^2}{b^2} &  0 & 0\\
0 & -\frac{b^2  + d^2 + \sqrt{x_2}}{2\,c^2} & 0\\
0 & 0 & -\frac{b^2  + d^2 - \sqrt{x_2}}{2\,c^2}
\end{array}\right]
\end{equation}  
 where     
  \begin{equation}
   x_2 = (b^2 + 2\, c^2)^2 + 2\, (b^2 - 2\,c^2)\, d^2 + d^4
  \end{equation}   
     
     \begin{equation}
D_3 = \left[\begin{array}{ccc}
\frac{a^2 + c^2 - d^2}{a^2} &  0 & 0\\
0 & -\frac{a^2  + d^2 +\sqrt{x_3}}{2\,c^2} & 0\\
0 & 0 & -\frac{a^2  + d^2 - \sqrt{x_3}}{2\,c^2}
\end{array}\right]
\end{equation}  
 where     
  \begin{equation}
   x_3 = (a^2 + 2\,c^2)^2 + 2\,( a^2 - 2\,c^2)\,d^2 + d^4
  \end{equation}   
      \begin{equation}
D_4 = \left[\begin{array}{ccc}
\frac{a^2 + b^2 - d^2}{a^2} &  0 & 0\\
0 & -\frac{a^2  + d^2 +\sqrt{x_4}}{2\,b^2} & 0\\
0 & 0 & -\frac{a^2  + d^2 -\sqrt{x_4}}{2\,b^2}
\end{array}\right]
\end{equation}  
 where     
  \begin{equation}
  x_4 = (a^2 + 2\,b^2)^2 + 2\,( a^2 - 2\,b^2)\,d^2 + d^4
  \end{equation}   
  Note that  $x_i > 0$ for $i=2,3,4$. 
  
If $\det(G_i)>0$ then all the eigenvalues of $A_i$ $(i=2,3,4)$ become negative and all the eigenvalues of $H_i$, for $i=2,3,4$  become real. Then $\det (H_i) = -\lambda :\lambda>0$ defines an ellipsoid.
   
If $\det(G_i)<0$ then only one eigenvalue of  $A_i$ $(i=2,3,4)$  is negative and in this case, $\det(H_i)= \lambda$, $\det(H_i)=  0$, $\det(H_i)=  -\lambda :\lambda>0 $ for $i=2,3,4$ define hyperboloid of one sheet, quadric cone and hyperboloid of two sheets respectively.
   
Finally, $A_1$ matrix is given as 
{
\renewcommand{\arraystretch}{2}
\begin{equation}
A_1 
= \left[
\begin{array}{ccc}
 -\frac{a^2 + b^2 + c^2}{c^2} & \frac{(a^2 + b^2)\,d}{a\,c^2} & \frac{b\,d}{a\,c}\\
\frac{(a^2 + b^2)\,d}{a\,c^2} & \frac{(a^2 + b^2)(c^2 - d^2)}{a^2\,c^2} & \frac{b\,(c^2 - d^2)}{a^2\,c} \\
\frac{b\,d}{a\,c} & \frac{b\,(c^2 - d^2)}{a^2\,c} & \frac{a^2 + c^2 - d^2}{a^2}
\end{array}
\right]
\end{equation}
}
  
Finding eigenvalues of $A_1$ is cumbersome. 
Therefore we proceed with a different approach here. 
  \begin{equation}
         \det (H_i) = v^T A_i v = \tilde{v}^T D_i \tilde{v}  =\lambda_1 (\tilde{k}_1)^2 + \lambda_2 (\tilde{k}_2)^2 + \lambda_3 (\tilde{k}_3)^2, \quad   \, v^T =(k_1, k_2, k_3); \quad \tilde{v}^T =(\tilde{k}_1, \tilde{k}_2, \tilde{k}_3)
     \end{equation}
     Where $D_i = P H P^T$ and $\tilde{v} = P^T v$. Now
     \begin{equation}
         \det(A_1)=\lambda_1 \lambda_2 \lambda_3 =-\frac{(det(G_1))^2}{a^2\,c^2  } < 0
     \end{equation}
This implies all $\lambda_i$ are negative or only one of them is negative. Now consider the following equations,
     \begin{equation}\label{eq:TrA1}
              Tr(A_1) = \lambda_1 + \lambda_2 + \lambda_3 = -\frac{a^4 + a^2\, b^2 + (d^2- c^2)(a^2 + b^2 + d^2 ) }{a^2\,c^2}
     \end{equation}
   and  
   \begin{equation}\label{eq:lambdailambdaj}
        \lambda_1\lambda_2 + \lambda_1\lambda_3+ \lambda_2\lambda_3 =det(G_1)\frac{2\,a^2 + b^2 + c^2 }{a^2\,c^2}
     \end{equation}
We will show that all eigenvalues of $A_1$ are negative if and only if $\det(G_1) > 0$. Thus when $\det(G_1) < 0$ only one eigenvalue can be negative.

From Eq.~(\ref{eq:TrA1})
\begin{equation}\label{eq:detG1}
    \det(G_1) >0 \implies Tr(A_1) < 0 
\end{equation}

Now assume that only one eigenvalue say ($\lambda_1$) is negative, then Eq.~(\ref{eq:TrA1}) and Eq.~(\ref{eq:lambdailambdaj}) imply  $\det(G_1) < 0 $ which contradict Eq.~(\ref{eq:detG1}). Thus  $\det(G_1) >0 $ implies all eigenvalues of $A_1$ are negative.

Similarly when all eigenvalues are negative Eq.~(\ref{eq:lambdailambdaj}) implies that $\det(G_1) >0 $,
implying further that $\det(G_1)>0$ iff all the eigenvalues of $A_1$ are negative. In this case,  $det (H_1) =-\lambda :\lambda>0$ defines an ellipsoid.

If $\det(G_1)<0$ then only one eigenvalue of  $A_1$ is negative and in this case $\det(H_1)= \lambda$, $\det(H_i)=  0$, $\det(H_i)= -\lambda :\lambda>0 $ and defines hyperboloid of one sheet, quadric cone, and hyperboloid of two sheets respectively.

From the above discussion, we have the following
important result:

\begin{thm}
Let $G\in M_{2}(\mathbb{C})$ be a fixed Hermitian matrix and let $H$ denote the ensemble of trace-less matrices satisfying $H^{\dagger} = 
GHG^{-1}$. When $\det(G)>0$,  $H$ has real eigenvalues and $det(H) = - \lambda : \lambda >0$ , defines an ellipsoid. Corresponding to all points on the ellipsoid, $H$ has unbroken PT-symmetry. When $\det(G)<0$, $H$ can have real or complex conjugate 
pairs of eigenvalues and $\det(H)= \lambda$, $\det(H)= 0$, $\det(H)=  -\lambda : \lambda >0 $ defines 
hyperboloid of one sheet, quadric cone, and hyperboloid of two sheets respectively. Corresponding to all the points on hyperboloid of one sheet, $H$ has broken PT-symmetry and corresponding to  all the points on quadric cone, hyperboloid of two sheets, $H$ has unbroken PT-symmetry.
\end{thm}

  \subsection{Examples}
  \begin{enumerate}
    \item For 
     \begin{equation}
         G = \left[\begin{array}{cc}
3  &  1-2i \\
1+2i  &  \quad 3
\end{array}\right] \in G_4
\end{equation} 

The general traceless G-Pseudo Hermitian matrix $H \in H_4$ for the given $G$ is 
\begin{equation}
H = \left[\begin{array}{cc}
 3 y + iz\quad & \quad \frac{x}{2}+y+\frac{3z}{2} -i(x+2y) \\
\frac{x}{2}-y+\frac{3z}{2} +i(x-2y)  &  -y - iz 
\end{array}\right]: x,y,z \in \mathbb{R}
\end{equation}
\begin{figure}[h]
    \centering
    \includegraphics[width=0.3\linewidth]{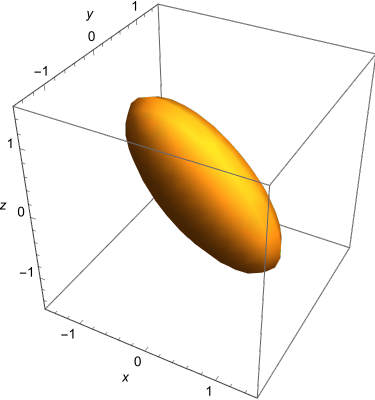}
    \caption{Contour Plot of  $det(H) =\tfrac{-5x^2}{4} -4y^2-\tfrac{5z^2}{4}- \tfrac{3xz}{2} =-1 $}
    \label{fig:1}
\end{figure}\\

    \item For 
     \begin{equation}
         G = \left[\begin{array}{cc}
\frac{1}{2}  &  -i \\
i  &  \quad \frac{1}{2}
\end{array}\right] \in G_6
     \end{equation} 

  \end{enumerate}

The general traceless G-Pseudo Hermitian matrix $F \in H_6$ for the given $G$ is 
\begin{equation}
F = \left[\begin{array}{cc}
 \frac{-y}{2} + iz\quad & \quad \frac{z}{2} +i(y -x) \\
\frac{z}{2}  +i(y + x)  & \frac{y}{2} - iz 
\end{array}\right]: x,y,z \in \mathbb{R}
\end{equation}

\begin{figure}[h]
    \centering
    \includegraphics[width = 0.85\linewidth]{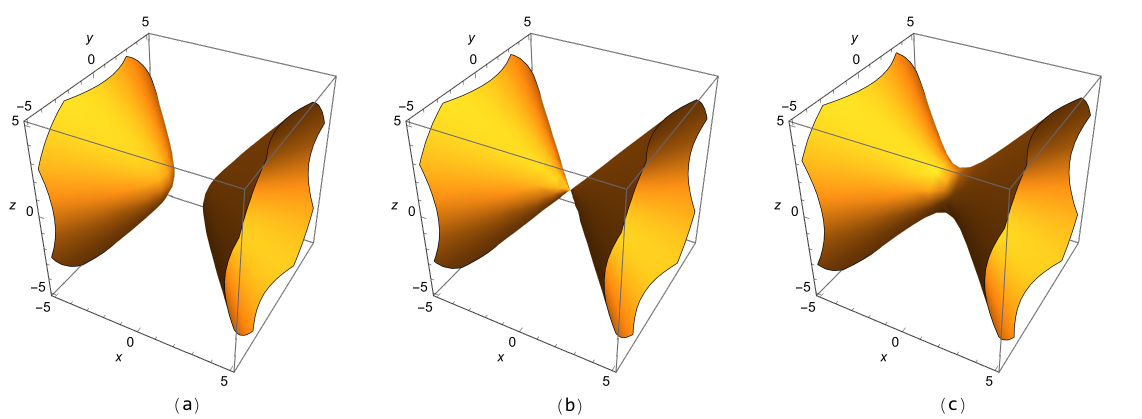}
    \caption{Contour Plots of $ det(F) =-x^2 + \frac{3y^2}{4} + \frac{3z^2}{4}$ for the case of (a) $det(F)=-1$, (b) $det(H_6)=0$, and (c)  $det(F)=1$.}
    \label{fig:2}
\end{figure}

\section{Determinant of General $G_s$-Pseudo Hermitian Matrices and Quadrics}\label{sec:6}
 In this section, we compute the determinant of the general traceless PT-symmetric matrices given by $H'_i (i=1...7)$ (refer to section~\ref{sec:4} for details) for singular $G_s$ and examine the associated quadrics.
\begin{equation}
\det (H'_i) = m^T A'_i m  : A'_i = {A'_i}^T \in M_3(\mathbb{R}), \, m^T =(m_1,m_2,m_3)
     \end{equation}
     Where $m_1,m_2,m_3$ are the free variables of the associated ensembles of matrices and $A'_i$ coincide with $A_i$ of $G-$pseudo Hermitian matrices when $d$ in $A_i$ equal to $ \pm \left|\pmb{g}_{R,i}\right| $ (refer section~\ref{sec:5}). The determinant of the ensembles  for each case is given below. 
     \begin{eqnarray}
        \det(H'_1) &=& -\frac{(d\,m_1 + b\,m_2 - a\,m_3)^2}{c^2}\\
        \det(H'_2) &=& -\frac{(m_1\,d + m_2\,b)^2}{c^2}\\
        \det(H'_3) &=& -\frac{(m_1\,d - m_2\,c)^2}{c^2}\\
        \det(H'_4) &=& -\frac{(m_1\,d + m_3\,a)^2}{b^2}\\  
        \det(H'_5) &=& -(m_1)^2\\  
        \det(H'_6) &=& -(m_1)^2\\
        \det(H'_7) &=& -(m_1)^2\\
  \end{eqnarray}
   In all case $H'_i$ has real eigenvalues, $\det (H'_i)=-1$ defines two parallel planes  and  $\det (H'_i)=0 $ defines a plane, for(i=1 to 7).
   
     From the above discussion, we have the following Theorem.
 \begin{thm}
Let $G\in M_{2}(\mathbb{C})$ be a  singular Hermitian matrix and let $H$ denote the ensemble of trace-less PT-symmetric matrices satisfying $H^{\dagger}G = 
GH$. Then $H$ has real eigenvalues and $det(H) = - \lambda$, $det(H) =0 : \lambda >0$ , defines two parallel planes and a single plane respectively. Corresponding to all points on the parallel planes and the single plane  $H$ has unbroken PT-symmetry.
\end{thm}
     
\subsection{Examples}
       \begin{enumerate}
    \item For 
   \begin{equation}
         G_s = \left[\begin{array}{cc}
\sqrt{5}  &  1-2i \\
1+2i  &  \quad \sqrt{5}
\end{array}\right] \in G_4
\end{equation} 

The general trace-less $G_s$-Pseudo Hermitian matrix $H'\in H'_4 $ for the given $G_s$ is 
~\vspace*{-0.5\baselineskip}
\begin{equation}
H' = \left[\begin{array}{cc}
 -\frac{y\sqrt{5} }{2} + iz \quad & \quad \frac{( 1-2i)(x-y)+z\sqrt{5}  }{2} \\
\frac{( 1+2i)(x+y)+z\sqrt{5}  }{2}& \quad  -\frac{ y\sqrt{5}}{2} - iz
\end{array}\right]: x,y,z \in \mathbb{R}
\end{equation}
\begin{figure}[h]
    \centering
    \includegraphics[width=0.75\linewidth]{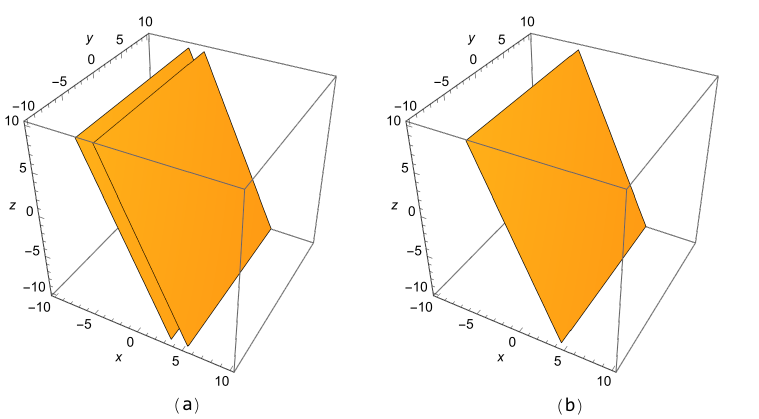}
    ~\vspace*{-1\baselineskip}
    \caption{Contour Plots of  $det(H') =-\tfrac{(x\sqrt{5} + z)^2}{4}$ for the case (a) $det(H')= -1$, (b) $det(H')= 0$.}
    \label{fig:3}
\end{figure}

\item 
For
~\vspace*{-1\baselineskip}
\begin{equation}
 G_s = \left[\begin{array}{cc}
1  &  -i \\
i  &  \quad 1
\end{array}\right] \in G_6
\end{equation} 
The general trace-less $G_s$-Pseudo Hermitian matrix $F'\in H'_6 $ for the given $G_s$ is 
~\vspace*{-0.5\baselineskip}
\begin{equation}
F' = \left[\begin{array}{cc}
-y  + iz\quad  & \quad i(y -x) + z \\
i(y + x) + z  & \ y - iz 
\end{array}\right]: x,y,z \in \mathbb{R}
\end{equation}
~\vspace*{-1\baselineskip}
\begin{figure}[h]
    \centering
    \includegraphics[width=0.75\linewidth]{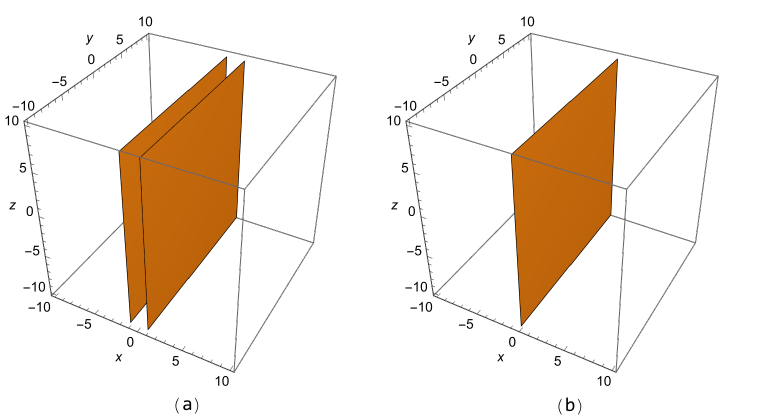}
    \caption{Contour Plots of  $det(F') =-x^2$ for the case (a) $det(F')= -1$, (b) $det(F')= 0$.}
    \label{fig:4}
\end{figure}
\end{enumerate}

\newpage
\section{Hermitian Matrices and Intersection of Quadrics}\label{sec:7}

In section~\ref{sec:4}  we have seen that all $G(G_s)$-pseudo Hermitian matrices for a given $G(G_s)$ can be found by solving a system of homogeneous equations given by Eq.~(\ref{eq:MX=0}).
Now for a given $G(G_s)$-Pseudo Hermitian matrix $H\in M_2(\mathbb{C})$, all $G(G_s)$  can be found by  rearranging Eq.~(\ref{eq:MX=0}) to get a system of quadratic equations. In this way, we show that there is a correspondence between the intersection of six quadrics and every Hermitian matrices with a fixed trace in general. From Eq.~(\ref{eq:hRgR2}) and Eq.~(\ref{eq:hIg0R2}) we get six quadrics of the following form and the intersection of those quadrics defines the entries of $G$($G_s$) for a given $H$ satisfying Eq.~(\ref{eq:HdaggerG}).
  \begin{equation}\label{eq:vtb1}
    \pmb{v}^T B_j \pmb{v} + \pmb{b}_j ^T \pmb{v} + c_j =0,\, j=1,2,3,4,5,6 
  \end{equation}
  where
\begin{equation}
    B_j = B_j^T \in\mathbb{M}_3(\mathbb{R}), \, \pmb{v}^T =(g^1_{R} = x, g^2_{R} = y, g^3_{R} = z), \, \pmb{b}_j^T =(b^{1}_{j}, b^{2}_{j}, b^{3}_{j}), \, c_j \in\mathbb{R},  
\end{equation}
\begin{eqnarray}
B_1 &=& \left[\begin{array}{ccc}
0 & \frac{-h^2_{R}}{2}  & \frac{-h^3_{R}}{2} \\
\frac{-h^2_{R}}{2}   &h^1_{R}& 0\\
\frac{-h^3_{R}}{2} & 0 &h^1_{R}
\end{array}\right] ;\, \pmb{b}_1^T= d(0,-h^3_{I},h^2_{I}) ;\, c_1 = 0
\\
B_2 &=& \left[\begin{array}{ccc}
h^2_{R} & \frac{-h^1_{R}}{2}  &0 \\
\frac{-h^1_{R}}{2}   &0&\frac{-h^3_{R}}{2} \\
0 &\frac{-h^3_{R}}{2}  &h^2_{R}
\end{array}\right];\, \pmb{b}_2^T= d(h^3_{I},0,-h^1_{I}) ;\, c_2 = 0
\\
B_3 &=& \left[\begin{array}{ccc}
h^3_{R} & 0 &\frac{-h^1_{R}}{2} \\
0   &h^3_{R}&\frac{-h^2_{R}}{2} \\
 \frac{-h^1_{R}}{2}&\frac{-h^2_{R}}{2}  &0
\end{array}\right];\, \pmb{b}_3^T=d(-h^2_{I},h^1_{I},0) ;\, c_3 = 0
\\
B_4 &=& \left[\begin{array}{ccc}
-h^1_{I} & \frac{-h^2_{I}}{2}  &\frac{-h^3_{I}}{2}  \\
 \frac{-h^2_{I}}{2}  &0&0\\
\frac{-h^3_{I}}{2} &0  &0
\end{array}\right];\, \pmb{b}_4^T= d(0,h^3_{R},-h^2_{R}) ;\, c_4 = h^1_{I}\, d^2
\\
B_5 &=& \left[\begin{array}{ccc}
0 & \frac{-h^1_{I}}{2}  &0 \\
\frac{-h^1_{I}}{2}    & -h^2_{I}&\frac{-h^3_{I}}{2} \\
0 & \frac{-h^3_{I}}{2} &0
\end{array}\right];\, \pmb{b}_5^T= d(-h^3_{R},0,h^1_{R}) ;\, c_5 = h^2_{I}\, d^2
\\
B_6 &=& \left[\begin{array}{ccc}
0 & 0 &\frac{-h^1_{I}}{2}  \\
0   &0&\frac{-h^2_{I}}{2} \\
\frac{-h^1_{I}}{2} & \frac{-h^2_{I}}{2} & -h^3_{I}
\end{array}\right];\, \pmb{b}_6^T= d(h^2_{R},-h^1_{R},0) ;\, c_6 = h^3_{I}\, d^2
\end{eqnarray} with $d = g^0_R$.
We analyze these six quadrics for the following two cases:

\vskip\baselineskip
\noindent
{\bf Case 1}: $d=0$

In this case Eq.~(\ref{eq:vtb1}) can be expressed in  standard form,
  \begin{equation}\label{eq:vtb2}
 \pmb{v}^T B_j \pmb{v}  = \tilde {\pmb{v}}^T D_j  \tilde {\pmb{v}}  =0 : \, \tilde {\pmb{v}}^T =(\tilde g^1_{R} = \tilde{x},\tilde g^2_{R} = \tilde{y}, \tilde g^3_{R} = \tilde{z}) ;j=1,2,3,4,5,6
  \end{equation}
  where $D_j = P B_j P^T$ and $ \tilde {\pmb{v}} = P^T \pmb{v}$
  \begin{eqnarray}
D_j = \left[\begin{array}{ccc}
h^j_{R} & 0 & 0 \\
0  &\frac{h^j_{R} - \left|\pmb{h}_R\right|}{2}& 0\\
0 & 0 &\frac{h^j_{R} + \left|\pmb{h}_R\right|}{2} \end{array}\right]  ; j=1,2,3
\end{eqnarray}

and  

\begin{eqnarray}
D_j = \left[\begin{array}{ccc}
0 & 0 & 0 \\
0  &-\frac{\left|\pmb{h}_I\right| + h^k_{I} }{2}& 0\\
0 & 0 & \frac{\left|\pmb{h}_I\right| - h^k_{I} }{2} \end{array}\right]  ;j=4,5,6
\end{eqnarray} 
where $k=1,2,3$ for $j=4,5,6$ respectively.\\ Now consider Eq.~(\ref{eq:vtb2}) for $j=1$, for the following cases.

1) When $h^1_{R}\neq0$ and $\left|h^1_{R}\right|\neq\left|\pmb{h}_R\right|$ then Eq.~(\ref{eq:vtb2}) defines a quadric cone.

2) When $h^1_{R} =0$ and $\left|\pmb{h}_R\right|\neq0$, Eq.~(\ref{eq:vtb2}) defines two planes passing through the origin.

3) When $\left|h^1_{R}\right|=\left|\pmb{h}_R\right|\neq0$, $B_1$ becomes diagonal and Eq.~(\ref{eq:vtb2}) defines a line.

\vskip\baselineskip
Exactly the same structures emerge from 
Eq.~(\ref{eq:vtb2}) in the cases of $j=2$ and $j=3$.
%Similar structures are observed for $j=2,3$. 
Along the same lines, consider Eq.~(\ref{eq:vtb2}) for $j=4$, for the following cases. 

1) When $h^1_{I}\neq0$ 
and $\left|h^1_{I}\right|\neq\left|\pmb{h}_I\right|$, then Eq.~(\ref{eq:vtb2})  defines two  planes passing through the origin.

2) When $h^1_{I} =0$ and $\left|\pmb{h}_I\right|\neq0$, Eq.~(\ref{eq:vtb2}) defines two planes passing through the origin.

3) When $\left|h^1_{I}\right|=\left|\pmb{h}_I\right|\neq0$, $B_4$ becomes diagonal matrix and Eq.~(\ref{eq:vtb2}) defines a plane.

\vskip\baselineskip
%Similar structure is 
%found for $j=5,6$. 
Exactly the same structures emerge from 
Eq.~(\ref{eq:vtb2}) in the cases of $j=5$ and $j=6$.
When $d=0$, $\pmb{h}_R\neq0$ and $\pmb{h}_I = 0$ (i.e. for $H \in S_1$), then out of the six equations that Eq.~(\ref{eq:vtb2}) represents, only 
the first three equations survive. Similarly when $d=0$, $\pmb{h}_R = 0$ and $\pmb{h}_I \neq 0$ (i.e. for $H \in S_2$), out of the six equations (Eq.~(\ref{eq:vtb2}), only the last three equations
are non-trivial. The matrix $G$ can be obtained from the intersection of these three quadrics.

Therefore, for a given PT-symmetric $H\in M_2(\mathbb{C})$  all trace-less $G$ with which it can be $G$ -Pseudo Hermitian can be obtained by the intersection of six quadrics from the previously mentioned possible structures.
\vskip\baselineskip
\noindent
{\bf Case 2:} $d\neq0$(In this case we find all $G$ s with a fixed $ d \neq0$)

In this case,  Eq.~(\ref{eq:vtb1}) can be expressed in standard form as 
\begin{equation}\label{eq:vtb3}
  \pmb{v}^T B_j \pmb{v} + \pmb{b}_j ^T \pmb{v} + c_j =   \tilde{\pmb {v}}^T D_j \tilde {\pmb {v}} + \tilde {\pmb{b}}_j ^T \tilde {\pmb{v}} + c_j =0 : \tilde {\pmb{b}}_j^T =(\tilde b_j^1,\tilde b_j^2,\tilde b_j^3) ,\, c_j \in\mathbb{R};\quad j=1,2,3,4,5,6
  \end{equation}
  where the first term of the above equation is same as Eq.~(\ref{eq:vtb2}), $\tilde{\pmb{b}}_j = P^T \pmb{b}_j$. Further, 
  for $j = 1, 2, 3$, we have:
  \begin{equation}
\left|\tilde b_j^1\right| =  \left|\frac{d \,(h^j_{R}\,h^j_{I} - \pmb{h}_R \cdot \pmb{h}_I)}{\sqrt{|\pmb{h}_{R}|^2 - (h^j_{R})^2}}\right|; \quad  \left|\tilde b_j^2\right| =  \left|\frac{d( \,\pmb{h}_{R}\,\times \pmb{h}_{I}) \cdot \pmb{e}_j}{\sqrt{2|\pmb{h}_{R}|(|\pmb{h}_{R}| + h^j_{R})}}\right|; \quad \left|\tilde b_j^3\right| =  \left|\frac{d( \,\pmb{h}_{R}\,\times \pmb{h}_{I}) \cdot \pmb{e}_j}{\sqrt{2|\pmb{h}_{R}|(|\pmb{h}_{R}| - h^j_{R})}}\right| %, 
%\, \mathrm{for} \, j=1,2,3
  \end{equation}
  On the other hand, for $j=4,5,6$ we have:
\begin{equation}
\left|\tilde b_j^1\right| =  \left|\frac{d \,(h^j_{R}\,h^j_{I} - \pmb{h}_R \cdot \pmb{h}_I)}{\sqrt{|\pmb{h}_{I}|^2 - (h^k_{I})^2}}\right|; \quad  \left|\tilde b_j^2\right| =  \left|\frac{d( \,\pmb{h}_{R}\,\times \pmb{h}_{I}) \cdot \pmb{e}_k}{\sqrt{2|\pmb{h}_{I}|(|\pmb{h}_{I}| + h^k_{I})}}\right|; \quad \left|\tilde b_j^3\right| =  \left|\frac{d( \,\pmb{h}_{R}\,\times \pmb{h}_{I}) \cdot \pmb{e}_k}{\sqrt{2|\pmb{h}_{I}|(|\pmb{h}_{I}| - h^k_{I})}}\right| %, \, \mathrm{for} \, j=4,5,6
  \end{equation}

where $k = 1, 2, 3 $  for $j= 4, 5, 6$ respectively and $\pmb{e}_1 = \pmb{i}, \pmb{e}_2 = \pmb{j}, \pmb{e}_3 = \pmb{k}  $.  
Now consider Eq.~(\ref{eq:vtb3}) for $j=1$, for the following cases.
  
     1) When $h^1_{R} \neq 0$ and $\left|h^1_{R}\right|\neq\left|\pmb{h}_R\right|$,  
     Eq.~(\ref{eq:vtb3}) can be expressed as 
  \begin{eqnarray}\label{eq:dneq01}
   \lambda^1_{j}\left( \tilde x +\frac{\tilde b_j^1}{2{\lambda^1_{j}} }\right)^2 + \lambda^2_{j} \left(\tilde y +\frac{\tilde b_j^2}{2\lambda^2_{j}}\right)^2 + \lambda^3_{j}\left( \tilde z +\frac{\tilde b_j^3}{2\lambda^3_{j} }\right)^2 -\delta_j =0 , \,\,\mathrm{for} \,\, j=1 
    \end{eqnarray}
    where 
     \begin{eqnarray}
       \delta_j= \frac{(\tilde b_j^1)^2}{ 4\lambda^1_{j}}+\frac{(\tilde b_j^2)^2}{4 \lambda^2_{j}}+\frac{(\tilde b_j^3)^2}{ 4\lambda^3_{j}}, \,\,\mathrm{for}\,\, j=1
  \end{eqnarray}
     $\delta_j$ may be zero, positive or negative and $\lambda^1_{j},\lambda^2_{j},\lambda^3_{j}$ are the first, second and third diagonal entries of $D_j $ (for $j=1$) and none of them are equal to zero. When $\delta = 0 $, 
 Eq.~(\ref{eq:dneq01}) defines a quadric cone. When $\delta $ and  $h^1_{R}$ both
     having the same (opposite) signs, Eq.~(\ref{eq:dneq01}) defines a hyperboloid of 1 sheet (2 sheets).
     
    2-a) When $h^1_{R} = 0 $, $\left|\pmb{h}_R\right|\neq0$ and $\tilde b_j^1 \neq0 $
    
   In this case Eq.~(\ref{eq:vtb3}) can be expressed as,
    \begin{eqnarray}\label{eq:dneq02}
  \lambda^2_{j} \left(\tilde y +\frac{\tilde b_j^2}{2\lambda^2_{j}}\right)^2 + \lambda^3_{j} \left( \tilde z +\frac{\tilde b_j^3}{2\lambda^3_{j} }\right)^2 + \tilde b_j^1 \tilde{x} =0 ,  \, \mathrm{for} \,j=1
    \end{eqnarray}
    
    where $\lambda^1_{j} = 0, \quad\lambda^3_{j} = -\lambda^2_{j}, \quad |\tilde b_j^2| = |\tilde b_j^3|$,
    and the Eq.~(\ref{eq:dneq02}) defines a hyperbolic paraboloid.
    
  2-b) When $h^1_{R} = 0 $, $\left|\pmb{h}_R\right|\neq0$ and $\tilde b_j^1 = 0 $
    
   In this case Eq.~(\ref{eq:vtb3}) can be expressed as,
   
    \begin{eqnarray}\label{eq:dneq02b}
  \lambda^2_{j} \left(\tilde y +\frac{\tilde b_j^2}{2\lambda^2_{j}}\right)^2 + \lambda^3_{j} \left( \tilde z +\frac{\tilde b_j^3}{2\lambda^3_{j} }\right)^2 =0, \, 
  \mathrm{for} \, j=1 
    \end{eqnarray} 

  The Eq.~(\ref{eq:dneq02b}) defines a pair of intersecting planes. 
  
  3)  When $\left|h^1_{R}\right| = \left|\pmb{h}_R\right|\neq0$ \\
    In this case,  $B_1$ becomes a diagonal matrix Eq.~(\ref{eq:vtb3}) can be expressed as
 
 \begin{eqnarray}\label{eq:dneq03}
  h^j_{R} \left( y +\frac{b_j^2}{2h^j_{R}}\right)^2 + 
   h^j_{R} \left(z +\frac{ b_j^3}{2h^j_{R} }\right)^2 -\delta_j =0 , for j=1
    \end{eqnarray}
  where  
     \begin{eqnarray}
       \delta_j=\frac{ (b_j^2)^2}{4 h^j_{R}}+ \frac{(b_j^3)^2}{ 4 h^j_{R}}, \,\, \mathrm{for}\,\,, j=1
  \end{eqnarray} 
Thus when $\delta_j\neq0$ Eq.~(\ref{eq:dneq03}) defines a cylinder and when $\delta_j = 0$ Eq.~(\ref{eq:dneq03}) defines a line.

4) When $\pmb{h}_R = 0$ and $\pmb{h}_I \neq 0$

In this case $B_1 = 0$ and Eq.~(\ref{eq:vtb3}) can be expressed as
\begin{eqnarray}\label{eq:dneq04}
    h^2_I z - h^3_I y = 0 
\end{eqnarray} 
Eq.~(\ref{eq:dneq04}) defines a plane.

A similar structure holds for $j=2,3$.\\

Along the same line Consider  Eq.~(\ref{eq:vtb3}) for $j=4$, for the following cases

1-a) When $h^1_{I} \neq0 $, $\left|h^1_{I}\right|\neq\left|\pmb{h}_I\right|$ and $\tilde{b}_j^1 \neq0$.

In this case Eq.~(\ref{eq:vtb3}) can be expressed as
 
        \begin{eqnarray}\label{eq:dneq05}
         \lambda^2_{j} \left(\tilde y +\frac{\tilde b_j^2}{2\lambda^2_{j}}\right)^2 + \lambda^3_{j}\left( \tilde z +\frac{\tilde b_j^3}{2\lambda^3_{j} }\right)^2 +\tilde b_j^1\left( \tilde x +\delta_j\right) =0 , \mathrm{for}\,\, j=4
    \end{eqnarray}
  where $\lambda^2_{j} < 0,~ \lambda^3_{j} > 0 $ are the second and third diagonal entries of $D_j $ (for $j=4$).
  \begin{eqnarray}
     \delta_j= \frac{1}{\tilde b_j^1} \left[c_j - \left(\frac{(\tilde b_j^2)^2}{4 \lambda^2_{j}}+ \frac{(\tilde b_j^3)^2}{ 4 \lambda^3_{j}}\right)\right], \,\, \mathrm{for}\,\, j=4
  \end{eqnarray} 
  Thus Eq.~(\ref{eq:dneq05}) defines a hyperbolic paraboloid.

1-b) When $h^1_{I} \neq0 $, $\left|h^1_{I}\right|\neq\left|\pmb{h}_I\right|$ and $\tilde{b}_j^1 = 0$.
In this case, Eq.~(\ref{eq:vtb3}) can be expressed as 

 \begin{eqnarray}\label{eq:dneq5b}
         \lambda^2_{j} (\tilde y  +\frac{\tilde b_j^2}{2\lambda^2_{j}})^2 + \lambda^3_{j}( \tilde z  +\frac{\tilde b_j^3}{2\lambda^3_{j} })^2  +\delta_j =0 ,for j=4
    \end{eqnarray}
    where  $ b_j^1 =0 $, and
    \begin{eqnarray}
\delta_j = c_j - \left(\frac{(\tilde b_j^2)^2}{4 \lambda^2_{j}}+ \frac{(\tilde b_j^3)^2}{ 4 \lambda^3_{j}}\right)
     ,for j=4 
  \end{eqnarray}
  Thus when $\delta_j\neq0$ Eq.~(\ref{eq:dneq5b}) defines a hyperbolic cylinder and when $\delta_j = 0$, Eq.~(\ref{eq:dneq5b}) defines two intersecting planes. 

2-a) When $h^1_{I} = 0 $, $\left|\pmb{h}_I\right|\neq0$, and $\tilde{b}_j^1 \neq0 $. In this case Eq.~(\ref{eq:vtb3}) can be expressed as 
  \begin{eqnarray}\label{eq:dneq06}
         \lambda^2_{j} \left(\tilde y +\frac{\tilde b_j^2}{2\lambda^2_{j}}\right)^2 + \lambda^3_{j}\left( \tilde z +\frac{\tilde b_j^3}{2\lambda^3_{j} }\right)^2 +\tilde b_j^1 \tilde x  =0 , \mathrm{for}\,\, j=4
    \end{eqnarray}
 where $\lambda^2_{j} = -\lambda^3_{j} $ and $|\tilde b_j^2| = |\tilde b_j^3|$. The Eq~.(\ref{eq:dneq06}) defines a hyperbolic paraboloid. 

2-b) When $h^1_{I} = 0 $, $\left|\pmb{h}_I\right|\neq0$, and $\tilde{b}_j^1 = 0$. In this case Eq.~(\ref{eq:vtb3}) can be expressed as 
 \begin{eqnarray}\label{eq:dneq06b}
  \lambda^2_{j} (\tilde y +\frac{\tilde b_j^2}{2\lambda^2_{j}})^2 + \lambda^3_{j}( \tilde z  +\frac{\tilde b_j^3}{2\lambda^3_{j} })^2 = 0 , for j=4
    \end{eqnarray}
   
  Thus Eq.~(\ref{eq:dneq06b}) defines two intersecting planes.

3) When $\left|h^1_{I}\right| = \left|\pmb{h}_I\right|\neq0$

In this case, $B_4$ becomes a diagonal matrix and Eq.~(\ref{eq:vtb3}) can be expressed as  
\begin{eqnarray}\label{eq:dneq07}
 -h^k_I x^2 + d b_j^2 y + d b_j^3 z +h^k_I  d^2 = 0, \quad for j=4
\end{eqnarray}

When both $b^2_j$ and $b^3_j$  become zero, Eq.~(\ref{eq:dneq07}) defines two parallel planes otherwise it defines a parabolic cylinder.

4) When  $\pmb{h}_I = 0$ and  $\pmb{h}_R \neq 0$  

In this case, Eq.~(\ref{eq:vtb3}) can be expressed as 

\begin{eqnarray}\label{eq:dneq08}
   b_j^2 y + b_j^3 z   =0
\end{eqnarray}
Eq.~(\ref{eq:dneq08}) defines a plane.

Similar structures are found for the cases of $j=5,6$ as well. From the above discussion we conclude that for a given PT-symmetric $H\in M_2(\mathbb{C})$, we get six quadrics among the above shown quadrics in general, such that thier intersection determine all the $G(G_s)$ matrices with which $H$ can be $G(G_s)$-pseudo Hermitian. Since all $2\times2$ PT-symmetric matrices are pseudo Hermitian we always get six quadrics in general, such that they intersect non trivially. 

\subsection{Examples}
\begin{enumerate}
    \item 
    For a given $H\in S_1$, all $G$ and $G_s$ with which it became pseudo hermitian can be obtained from proposition~\ref{prop:HS1}. The same can be also obtained from the intersections of quadrics defined by Eq.~(\ref{eq:vtb1}). For example consider a $H\in S_1$  such that $\pmb{h}_R=(1,1,1)$, then $G$ is given by
    \begin{equation}
 G =  \sigma_{0}\,d + \pmb{\sigma} \cdot \left (x, y, z\right)
\end{equation}
where $ (x,y,z)
=\lambda(1, 1, 1), :\lambda\in \mathbb{R}$. When $d=0$, it follows from Eq.~(\ref{eq:vtb1}) that the points $(x, y, z)$ come as the intersection of three quadric cones. Similarly, when $d = 1$ the points (x, y, z) come as the intersection of the same three quadric cones and three planes as shown in  Fig.~\ref{fig:5} and Fig.~\ref{fig:6}

\begin{figure}[h]
\centering
\includegraphics[width=.65\linewidth]{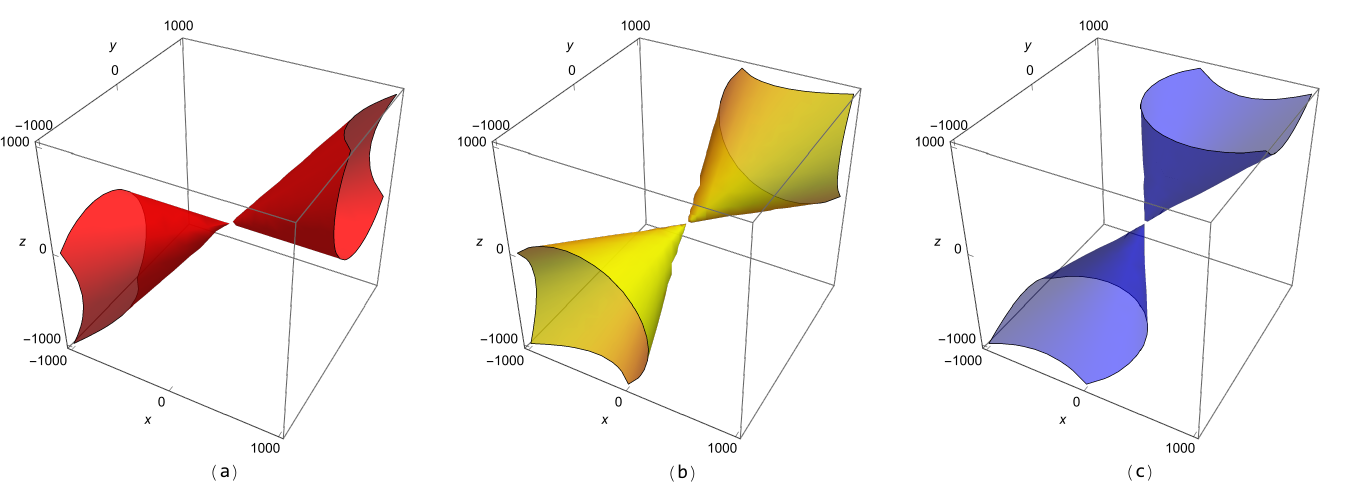}
\centering
\includegraphics[width=.65\linewidth]{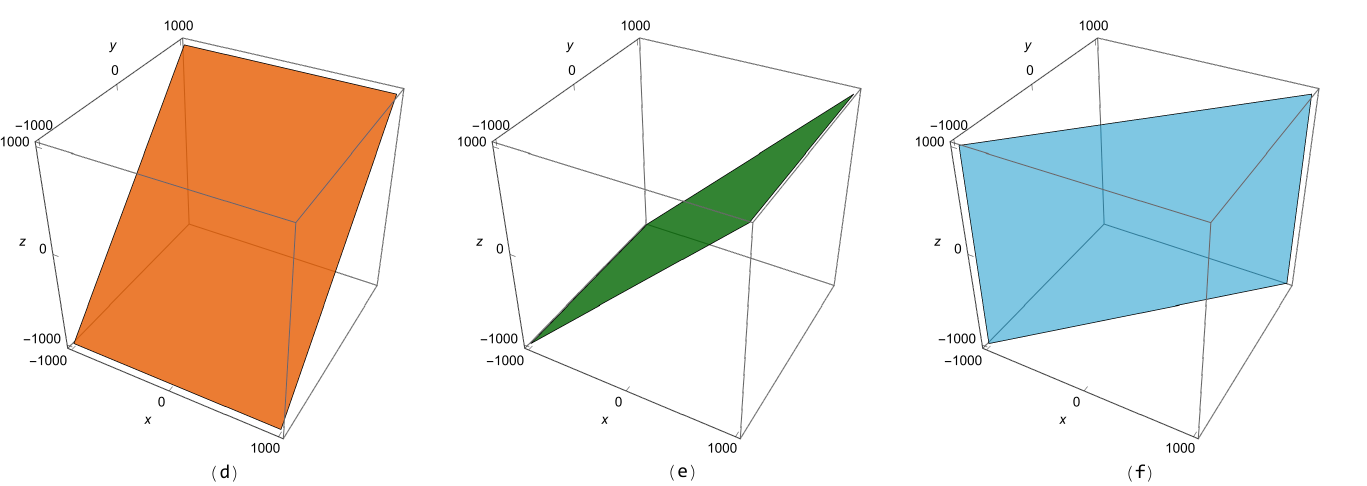}
\vspace*{-1\baselineskip}
\caption{Contour Plots of the quadrics, when   $\pmb{h}_R=(1,1,1)$, and  $\pmb{h}_I=(0,0,0)$.  (a) $y^2 + z^2 -x( y + z) = 0$, (b) $ x^2 + z^2 -y( x +z)  = 0$,  (c) $ x^2 + y^2 - z( x + y) =0$, (d) y - z = 0, (e) -x + z = 0, (f) x - y = 0.}
\label{fig:5}
\end{figure}
~\vspace*{-3\baselineskip}
\begin{figure}[h]
\centering
\includegraphics[width=0.65\linewidth]{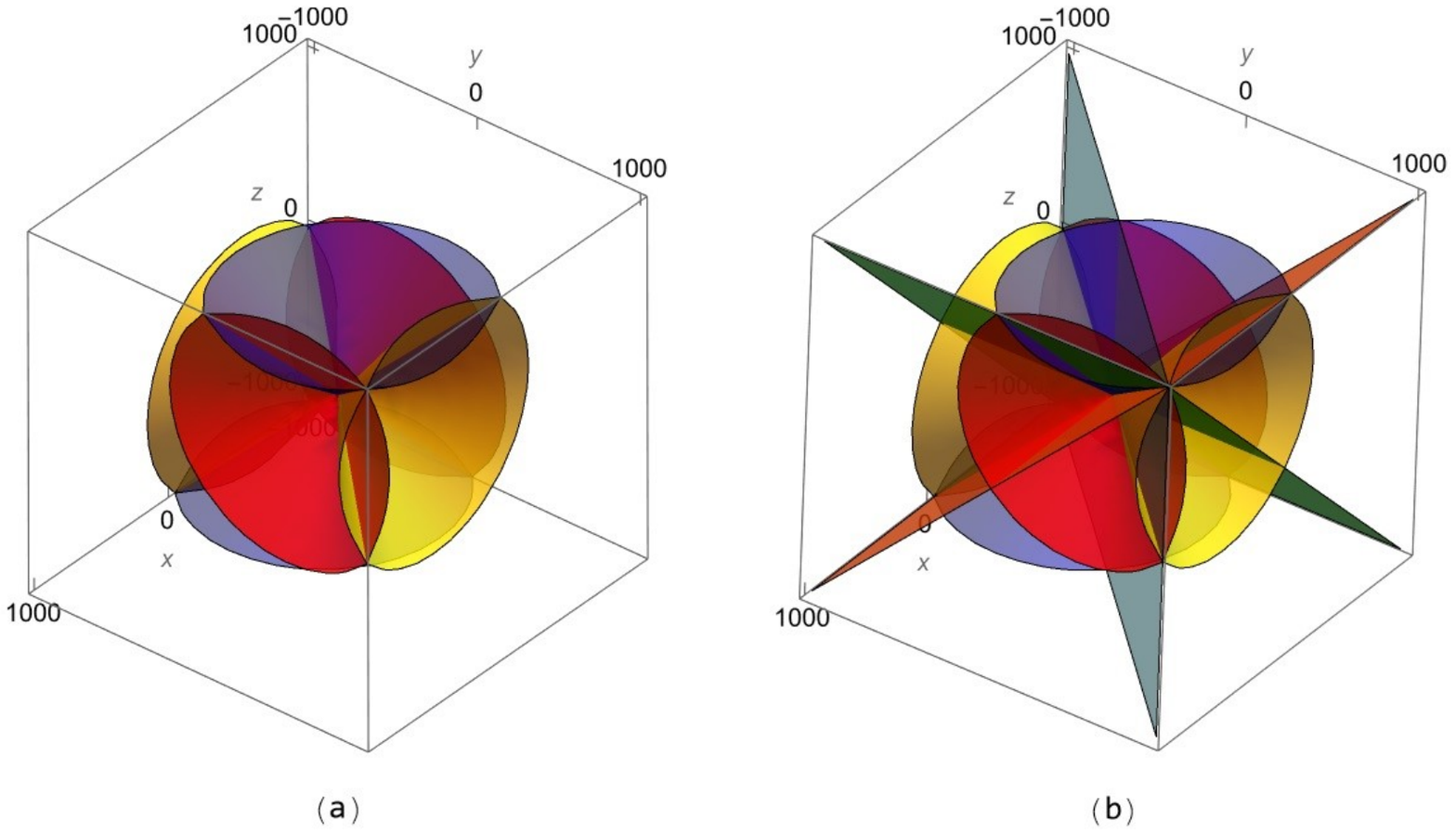}
\vspace*{-4\baselineskip}
\caption{Contour Plots of intersection of quadrics, when  $\pmb{h}_R=(1,1,1)$, and $\pmb{h}_I=(0,0,0)$, for the case (a) $d= 0$, (b) $d=1$}
    \label{fig:6}
\end{figure}

\newpage
\item    For a given $H\in S_2$, all $G$  with which it became pseudo hermitian can be obtained from proposition~\ref{prop:HS2} and Eq.~(\ref{eq:hI.gR=0}). The same can be also obtained from the intersections of quadrics defined by Eq.~(\ref{eq:vtb1}). For example consider a $H\in S_2$  such that $\pmb{h}_I=(1,1,1)$, then $G$ is given by
    
\begin{equation}
 G =   \pmb{\sigma} \cdot \left (x, y, z\right)
\end{equation}
where$ (x,y,z)
=(\lambda_1, \lambda_2, -(\lambda_1  + \lambda_2 )), :\lambda_1, \lambda_2 \in \mathbb{R}$. It follows from Eq.~(\ref{eq:vtb1}) that these points $(x, y, z)$ come as the intersection of three pairs of planes passing through the origin as shown in Fig.~\ref{fig:7}.

\begin{figure}[h]
    \centering
    \includegraphics[width=0.75\linewidth]{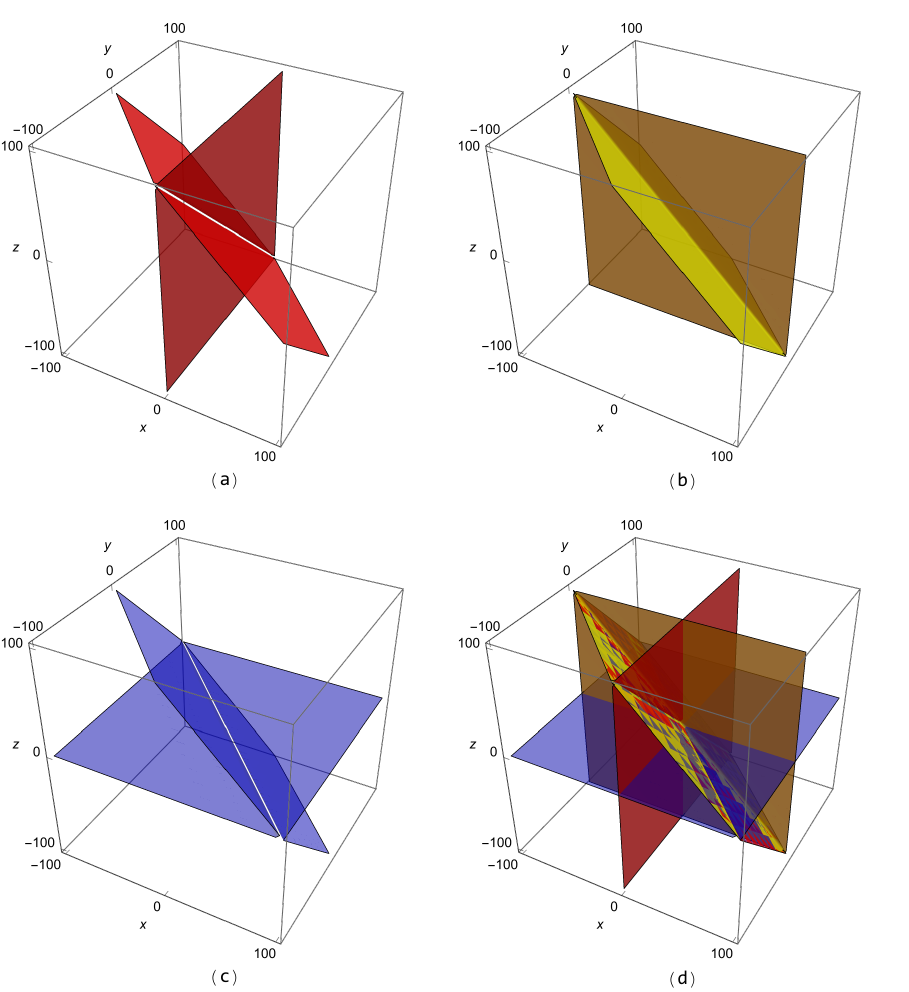}
    \caption{Contour Plots of quadrics, when  $\pmb{h}_R=(1,1,1)$, and $\pmb{h}_I=(0,0,0)$. (a) $x^2 + x( y +  z) = 0$, (b) $y^2 + y(x + z) = 0 $, (c) $z^2 +  z(x + y) = 0$, (d) intersection of the quadrics.}
    \label{fig:7}
\end{figure}

 \begin{figure}[h]
    \centering
    \includegraphics[width=.7\linewidth]{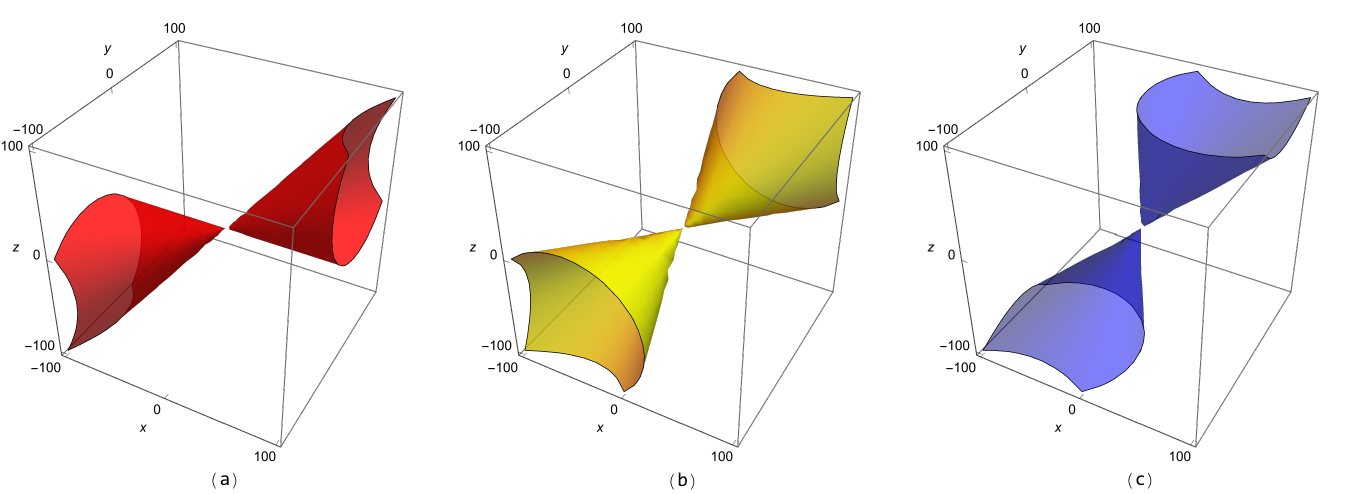}
        \centering
    \includegraphics[width=.7\linewidth]{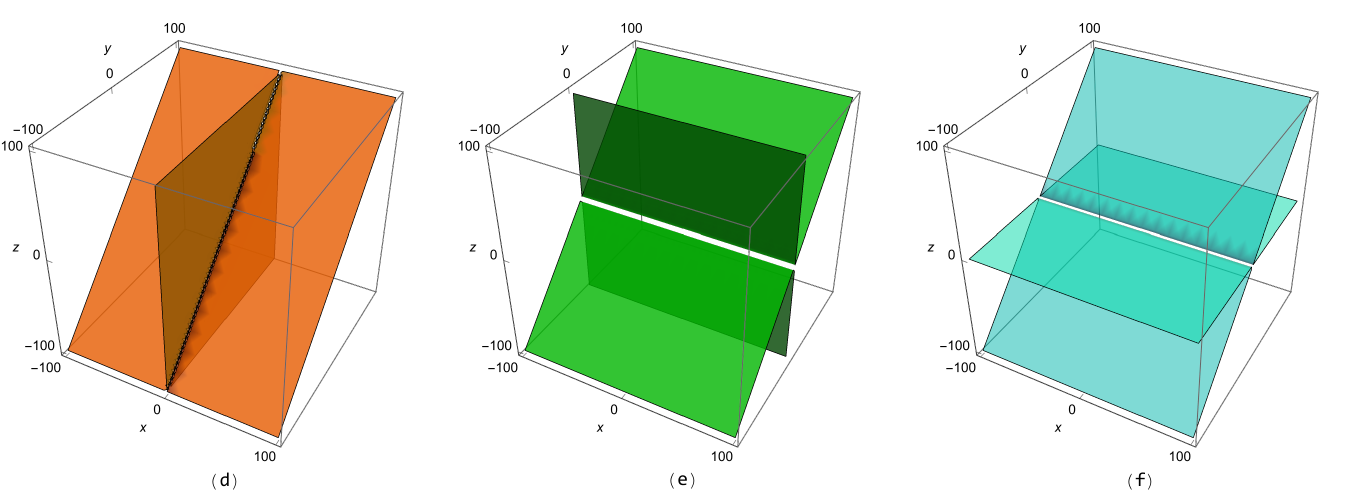}
    \caption{Contour Plots of quadrics, for  $\pmb{h}_R=(1,1,1)$, $\pmb{h}_I=(0, -1, 1)$ and $d=0$. (a) $ y^2 + z^2 - x( y +  z) = 0$, (b) $ x^2 + z^2 - y(x +  z) = 0 $, (c) $ x^2 + y^2 -  z(x +  y) = 0$, (d) $x(y - z) = 0$,(e) $y(y-  z) = 0$,(f) $z( y-z) =0 $}
    \label{fig:8}
~\vspace*{-1\baselineskip}
\end{figure}
\begin{figure}[h]
    \centering
\includegraphics[width=0.45\linewidth]{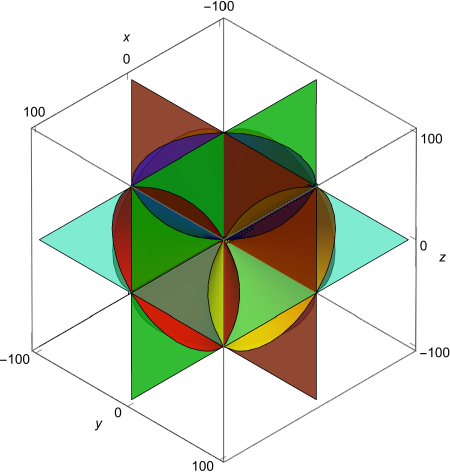}
    \caption{Contour Plot of intersection of quadrics, for  $\pmb{h}_R=(1,1,1)$, $\pmb{h}_I=(0,-1, 1)$, and $d=0$.}
    \label{fig:9}
\end{figure}

\begin{figure}[h]
    \centering
    \includegraphics[width=0.7\linewidth]{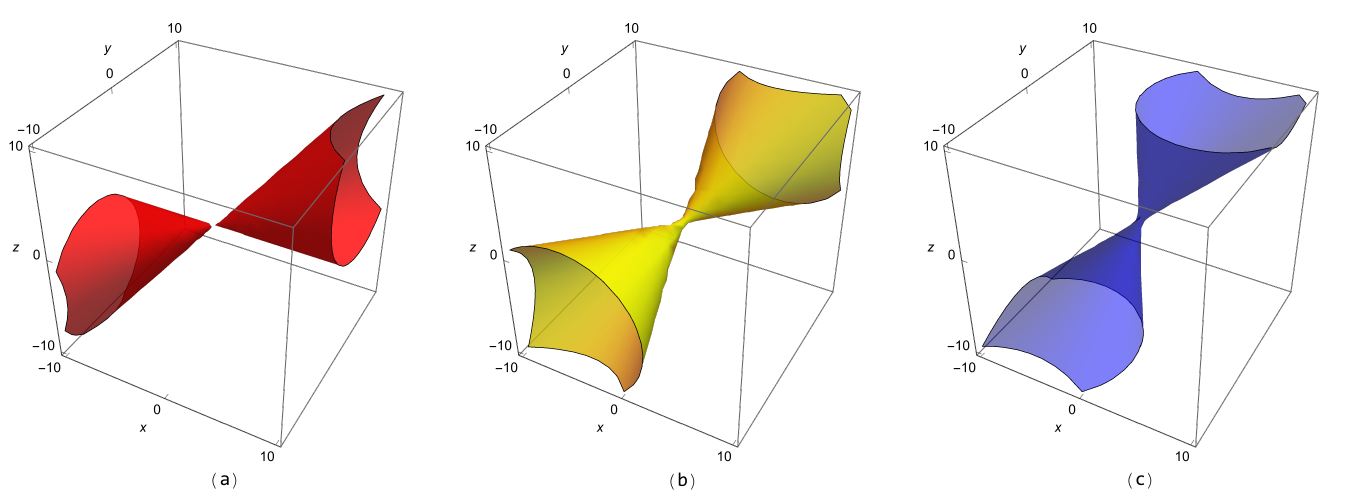}
        \centering
    \includegraphics[width=0.7\linewidth]{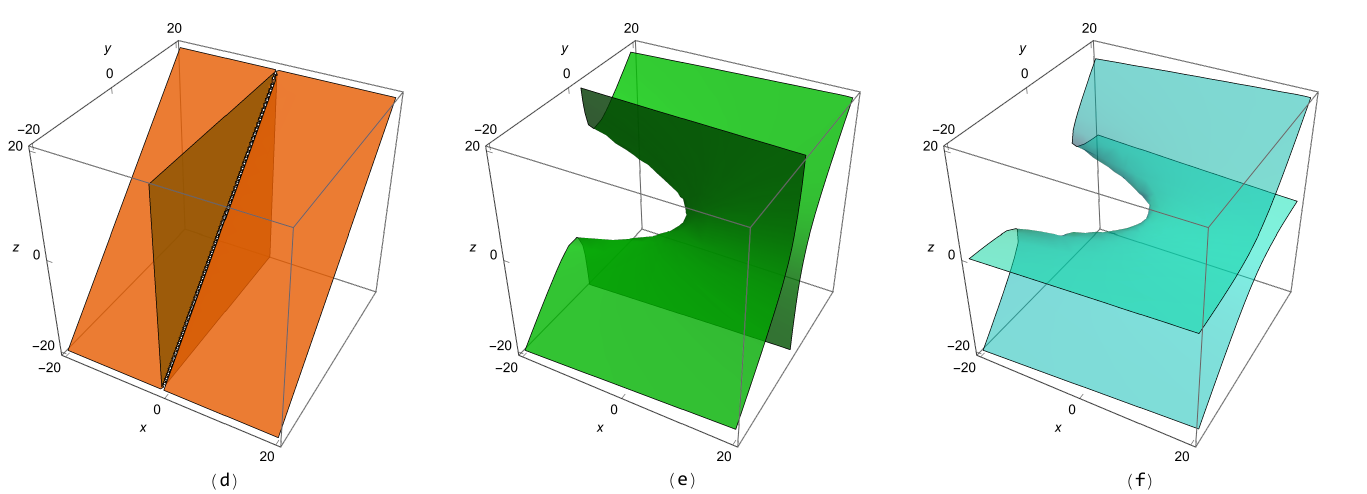}
    \caption{Contour Plots of quadrics, for  $\pmb{h}_R=(1,1,1)$, $\pmb{h}_I=(0, -1, 1)$ and $d=1$. (a) $ y^2 + z^2 - x( y +  z) -y - z  = 0$, (b) $ x^2 + z^2 - y(x +  z) + x= 0 $, (c) $ x^2 + y^2 - z(x + y ) + x = 0$, (d) $x(y - z) + y - z  = 0$, (e) $y(y- z) - x + z - 1= 0$, (f) $z(y-z) + x - y + 1 =0 $}
    \label{fig:10}
~\vspace*{-1\baselineskip}
\end{figure}
\begin{figure}[h]
    \centering
    \includegraphics[width=0.45\linewidth]{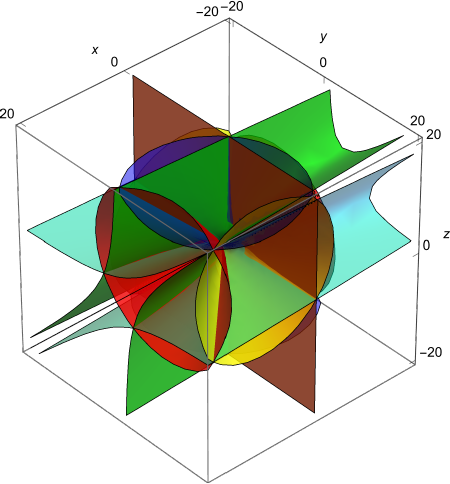}
    \caption{Contour Plot of intersection of quadrics, for  $\pmb{h}_R=(1,1,1)$, $\pmb{h}_I=(0,-1, 1)$, and $d=1$.}
    \label{fig:11}
\end{figure}

\newpage
 \item  For a given $H\in S_4$, from proposition~\ref{prop:HS4},  all trace less $G$ matrix with which $H$ can be Pseudo hermitian will be such that $\pmb{g}_R \times \pmb{h}_R = 0$. Similarly  all $G$ and $G_s$ for a given $d\neq0$ will be such that $\pmb{g}_R \cdot \pmb{h}_I = 0$ and $d= \frac{(\pmb{h}_R \times \pmb{g}_R) \cdot \pmb{h}_I}{\pmb{h}_I \cdot \pmb{h}_I}$. For example  consider a $H\in S_4$  with $\pmb{h}_R = (1,1,1)$ and $\pmb{h}_I = (0,-1,1)$. In this case the points $ (x,y,z)
=\lambda(1, 1, 1),:\lambda\in \mathbb{R}$  define $G$ when $d=0$, which can be obtained by the intersection of six quadrics defined by Eq.~(\ref{eq:vtb1}), where first three are quadric cones and last three are pairs of planes passing through the origin as shown in Fig.~\ref{fig:8} and Fig~\ref{fig:9}. When $d\neq0$ the points $ (x,y,z)
=(\lambda, \lambda + d,\lambda + d ), :\lambda\in \mathbb{R}$ define $G(G_s)$, where the points  $\lambda = \frac{(-2\pm1)d}{3}$ are corresponding to $G_s$.
These points (for d=1) can be obtained as the intersection 
of one quadric cone, two hyperboloid of  1 sheet, a pair of planes passing through the origin, and two hyperbolic paraboloid as shown in Fig.~\ref{fig:10} and Fig.~\ref{fig:11}.
\end{enumerate}

\phantomsection
~\newpage
~\newpage
~\newpage

\section{Summary}\label{sec:8}
Seven distinct ensembles of $G(G_s)$-Pseudo Hermitian matrices, which represent all possible $G(G_s)$-pseudo Hermitian matrices belonging to $M_2(\mathbb{C})$ have been studied systematically as a collection. We proved that all
 $2\times2$  $G$-Pseudo Hermitian matrices are PT-symmetric. All $H\in M_2(\mathbb{C})$  satisfying the relation $ H^{\dag}G = GH $, for a given Hermitian matrix $G(G_s)$ are found to be expressible as a linear variety . We found that for any two Hermitian $G_i,G_j\in M_2(\mathbb{C})$ such that $G_i\neq \lambda G_j$, there always exists exactly one trace less PT-symmetric $H\in M_2(\mathbb{C})$ (up to real scaling) which is pseudo-Hermitian with respect to both these $G$ matrices.  We have shown that the determinant of the ensemble of trace less $G(G_s)$-Pseudo Hermitian matrices exhibit quadric structures such that when $ \det(G) >0 $, one obtains ellipsoid which  represent  unbroken PT-symmetric Hamiltonians. When $ \det(G) <0 $, the quadrics can be  hyperboloid of 1-sheet, quadric cone or hyperboloid of two sheets such that  hyperboloid of 1 sheet represent broken  PT-symmetric Hamiltonians meanwhile hyperboloid of 2-sheets and quadric cone represent unbroken PT-symmetric Hamiltonians. In the case of $G_s$-Pseudo Hermitian matrices the quadrics associated with determinant  can be two parallel planes or a single plane and  these two cases represent  unbroken PT-symmetric Hamiltonians. 
 The set of all the matrices $G\in M_2(\mathbb{C})$, satisfying $H^{\dagger}G = GH$, given a specific $H\in M_2(\mathbb{C})$, are shown to be describable in terms of quadratic variety. Thus, for a 
 given PT-symmetric $H\in M_2(\mathbb{C})$, all the $G(G_s)$  with a fixed trace
 can be obtained from the intersections of six quadrics. The possible quadrics are found to be among the quadrics: quadric cone, hyperboloid of 1 sheet, hyperboloid of 2-sheets, hyperbolic paraboloid, cylinder, hyperbolic cylinder, parabolic cylinder, two planes passing through the origin, two parallel planes, a single plane and a straight line. In this way for any PT-symmetric $H\in M_2(\mathbb{C})$, we get six quadrics such that they intersect non trivially.
\section*{Acknowledgments}\label{sec:9}
S.A. acknowledges financial support from AFL Pvt. Ltd. through the 
Cyrus Guzder Fellowship.
We would like to express our sincere gratitude to Arnab Gowsami and Dr. Sudhir R. Jain for helpful discussions.

\section*{References}
\bibliographystyle{unsrt}
\bibliography{aipsamp}
\end{document}